\documentclass[a4paper,12pt]{article}
\pdfoutput=1
\usepackage{jheppub}
\usepackage{slashed}
\usepackage[dvipsnames,table]{xcolor}
\usepackage{hyperref}
\usepackage{xspace}
\usepackage[tight]{subfigure}
\usepackage{amsmath}
\usepackage{amssymb}
\usepackage{amsfonts}
\usepackage{mathrsfs}
\usepackage{comment}
\usepackage{afterpage}
\usepackage{verbatim}
\usepackage{booktabs}
\usepackage{array}
\usepackage[title,titletoc]{appendix}
\usepackage{tabularx}
\usepackage{physics}
\usepackage[T1]{fontenc}
\usepackage[normalem]{ulem}
\allowdisplaybreaks[4]

\newcommand{\nn}{\nonumber}

\newcommand{\pt}{p_T}
\newcommand{\y}{y}

\DeclareRobustCommand{\Sec}[1]{Sec.~\ref{sec:#1}}

\definecolor{darkgreen}{rgb}{0.13,0.55,0.13}

\title{Small radius inclusive jet production at the LHC through NNLO+NNLL}

\author[a]{Terry Generet,}
\affiliation[a]{Cavendish Laboratory, University of Cambridge, Cambridge CB3 0HE, UK}
\emailAdd{generet@hep.phy.cam.ac.uk}

\author[b]{Kyle Lee,}
\affiliation[b]{Center for Theoretical Physics, Massachusetts Institute of Technology, Cambridge, MA 02139, USA}
\emailAdd{kylel@mit.edu}

\author[c]{Ian Moult,}
\affiliation[c]{Department of Physics, Yale University, New Haven, CT 06511, USA}
\emailAdd{ian.moult@yale.edu}

\author[d]{Rene Poncelet,}
\affiliation[d]{Institute of Nuclear Physics, ul. Radzikowskiego 152, 31--342 Krakow, Poland}
\emailAdd{rene.poncelet@ifj.edu.pl}

\author[e]{Xiaoyuan Zhang}
\affiliation[e]{Department of Physics, Harvard University, Cambridge, MA 02138, USA}
\emailAdd{xiaoyuanzhang@g.harvard.edu}

\abstract{The study of hadronic jets and their substructure at hadronic colliders is crucial for improving our understanding of QCD, and searching for new physics. As such, there has been a significant effort to improve their theoretical description. In the small radius limit, inclusive jet production exhibits a universal factorization, enabling the resummation of logarithms which greatly stabilizes theoretical predictions. 
In this paper, we show how to combine a recently introduced framework for small-$R$ resummation with the \textsc{Stripper} subtraction formalism for fragmentation, enabling next-to-next-to-leading order calculations of small-$R$ inclusive jet production for a wide variety of processes at the LHC.
We extract the two-loop constants for the jet functions, enabling for the first time next-to-next-to-leading logarithmic resummation matched to next-to-next-to-leading order perturbative calculation.
We compare with CMS data for small-$R$ jet production, and find that our results greatly improve the accuracy of the predictions at small-$R$, and stabilize the perturbative convergence and error estimates at larger $R$.
Our approach is applicable to a wide class of jet substructure observables exhibiting similar factorization theorems, opening the door to an NNLO jet substructure program at the LHC.
}


\preprint{Cavendish-HEP-25/02, MIT-CTP 5857, IFJPAN-IV-2025-7}

\begin{document}
\strut\hfill
\maketitle

\section{Introduction}\label{sec:intro}
The study of jet production cross sections at high-energy hadron-hadron colliders constitutes an essential test of Quantum Chromodynamics (QCD). As such, there exists a plethora of precision jet measurements by the Large Hadron Collider's experimental collaborations ALICE \cite{ALICE:2013yva, ALICE:2019qyj}, ATLAS \cite{ATLAS:2011qvj, ATLAS:2013pbc, ATLAS:2014qmg, ATLAS:2014riz, ATLAS:2015yaa, ATLAS:2017kux, ATLAS:2017bje, ATLAS:2017ble,  ATLAS:2018sjf, ATLAS:2020vup, ATLAS:2023tgo, ATLAS:2024png}, and CMS \cite{CMS:2013vbb, CMS:2014mna, CMS:2014nvq, CMS:2016uxf, CMS:2016lna, CMS:2016jip, CMS:2020caw, CMS:2021yzl, CMS:2024hwr}.  These measurements allow for precision tests of perturbative QCD and the extraction of fundamental parameters such as the strong coupling constant $\alpha_s$ \cite{Giele:1995kb, Britzger:2017maj, ATLAS:2023tgo, CMS:2024rkg, Ahmadova:2024emn}. Moreover, jet phenomenology is central to the quest for higher precision in measurements at hadron colliders, like the Large Hadron Collider (LHC), since it touches on various aspects of precision phenomenology: tuning of Monte Carlo (MC) generators, extraction of parton distribution functions (PDFs) \cite{Nocera:2017zge, Harland-Lang:2017ytb, AbdulKhalek:2020jut}, and the modeling of backgrounds in many new-physics searches \cite{Lindert:2017olm, Llorente:2024wpr}.

Many of these tasks depend on the precision of the theoretical predictions and an accurate understanding of the associated uncertainties. The primary sources of uncertainty for high transverse momentum anti-$k_T$ jets~\cite{Cacciari:2008gp,Cacciari:2011ma} are the modeling of hadronization and the underlying event, and higher order perturbative corrections. The dependence of hadronization and the underlying event scale with the jet radius as $1/R$ and $R^2$, respectively,  allowing them to be disentangled and providing tests of the robustness of theoretical predictions \cite{Dasgupta:2007wa}. For small radii, the hadronization effects are enhanced, while for large radii the underlying event modeling plays a more crucial role, at intermediate ranges ($R=0.4-0.7$) the perturbative effects dominate. Fixed-order computations for jet production processes are available for many final states through next-to-next-to-leading order (NNLO) in QCD: boson production with jet \cite{Boughezal:2015dva, Boughezal:2015dra, Boughezal:2015aha, Chen:2014gva, Gehrmann-DeRidder:2015wbt, Boughezal:2015ded}, photon(s) with jet \cite{Campbell:2016lzl, Chawdhry:2021hkp, Badger:2023mgf}, inclusive jet and di-jet \cite{Currie:2016bfm, Currie:2017eqf, Currie:2018xkj, Czakon:2019tmo, Chen:2022tpk}, as well as three jet \cite{Czakon:2021mjy, Alvarez:2023fhi} final states. For smaller radii, fixed-order predictions becoming exceedingly unreliable due to logarithmic enhancements in the jet radius parameter. To restore the predictivity of perturbative QCD, the resummation of small-$R$ logarithms needs to be performed. Such resummed computations have been performed through NLO+NLL \cite{Dasgupta:2014yra, Kang:2016mcy, Dai:2016hzf, Dasgupta:2016bnd} and combined \cite{Liu:2017pbb, Liu:2018ktv,Moch:2018hgy} with NLL threshold resummation \cite{Kidonakis:1998bk, Kidonakis:2000gi}.

In addition to the calculation (or modeling) of the dominant effects, it is important to understand the uncertainties to make precise and accurate phenomenological statements. Keeping the uncertainties on non-perturbative effects aside (they are typically determined by discrete model variations), the uncertainties related to the finite perturbative expansion are notoriously difficult to estimate. Typically, uncertainties are estimated with the established method of scale variation.\footnote{Alternatives to scale variations to estimate uncertainties are discussed in Ref.~\cite{Cacciari:2011ze, Bonvini:2020xeo, Duhr:2021mfd, Ghosh:2022lrf, Tackmann:2024kci, Lim:2024nsk}.} In case of inclusive jet production, the pattern of perturbative convergence in fixed-order computations is particularly sensitive to the choice of the central scale \cite{Currie:2017eqf}. Furthermore, it has been observed that for intermediate jet radii ($R \sim 0.4$–$0.7$), perturbative corrections can either become significantly larger than the estimated uncertainties or lead to nearly vanishing uncertainties at NNLO QCD \cite{Bellm:2019yyh}. This gives another independent motivation to include small-$R$ resummation to remove some of the accidental cancellations that occur at fixed order and provide a more robust uncertainty estimate from the variation of scales.

Recently, an all-order factorization theorem for small radius jet production was presented in Ref.~\cite{Lee:2024icn}, and was illustrated at NLL~\cite{Lee:2024tzc} in the simple context of $e^+e^-$ colliders. The new factorization presented for single-inclusive jet processes in Ref.~\cite{Lee:2024icn} accounts for previously neglected logarithmic contributions starting at NLL accuracy and factorizes into a jet function, encoding the jet algorithm dynamics, and a hard function identical to that of single-inclusive hadron production. An alternative approach to the resummation of small-$R$ logarithms has been developed in Refs.~\cite{Dasgupta:2014yra,Dasgupta:2016bnd,vanBeekveld:2024jnx,vanBeekveld:2024qxs}, and illustrated at NLL. Since calculations for inclusive jet production are already performed at NNLO, improving the precision of these calculations requires small-$R$ resummation at NNLL, which has not previously been achieved. Beyond the factorization theorem, this requires the calculation of the two-loop jet function constants, as well as the NNLO hard functions for inclusive hadron production for hadron collider processes. Ultimately, one would like to obtain these in an automated way for generic processes at the LHC. 

The aim of the present work is twofold. First, we present a framework to achieve accurate predictions through NNLO QCD matched to NNLL resummation for the jet-radius parameter based on the computation of the inclusive jet function and the extension of the \textsc{Stripper} framework~\cite{Czakon:2010td, Czakon:2014oma, Czakon:2019tmo,Czakon:2021ohs,Czakon:2024tjr} to compute the necessary jet hard functions and convolutions. This provides a seamless interface between analytic factorization theorems based on single inclusive hard functions, and the \textsc{Stripper} framework. Using this, we are also able to extract the two-loop jet function constants. Second, we present a phenomenological analysis of NNLO+NNLL accurate jet spectra for various jet radii and a comprehensive comparison with the LHC data measured by CMS. We find that the inclusion of higher order corrections greatly improves the description of the data, as well as the robustness of uncertainty estimates from scale variation.

While the focus of this paper is on single inclusive jet production, our broader goal is to develop a framework to enable precision calculations of jet substructure observables at NNLO+NNLL for hadron colliders. The approach of this paper extends to any jet substructure observable that exhibits a factorization theorem onto a single inclusive hard function. This includes in particular energy correlator observables computed inside high energy jets \cite{Chen:2020vvp,Komiske:2022enw,Lee:2022ige,Chen:2023zlx}. We hope to report on the calculation of this exciting class of observables in future work.

An outline of this paper is as follows. In \Sec{framework}, we describe in detail the theoretical framework we have developed for our predictions. We review the factorization theorem for small-$R$ jet production, the subtraction formalism implemented in \textsc{Stripper} for fragmentation, and how these can be efficiently combined to achieve precision predictions of inclusive jet production at hadron colliders. We also describe our extraction of the two-loop constants for the jet functions appearing in the factorization theorem. In \Sec{phenomenology}, we perform a phenomenological study, comparing our state of the art predictions with CMS data. We study in detail the perturbative convergence, effects of resummation, and incorporate non-perturbative power corrections. We conclude and discuss future directions in \Sec{conc}.

\section{Framework}\label{sec:framework}

In this section, we describe the setup for the inclusive jet production in the small radius limit at hadron colliders. 

\subsection{Collinear factorization for small radius jet production}\label{subsec:factorization}

In small-radius jets, one can identify a hard scale associated with the jet transverse momentum $\sim \pt$ and a jet scale $\sim \pt R$. The factorization theorem separates dynamics at these scales and allows resummation of logarithms $\sum_{m\leq n} \alpha_s^n \ln^m R$ that are important for small-$R$ jets. At leading power (LP), such factorization for the differential cross section with respect to jet $\pt$ and rapidity $\eta$ are given as~\cite{Lee:2024icn,Lee:2024tzc}
\begin{align}\label{eq:factorization}
    \frac{d\sigma_{\rm LP}}{d\pt d\eta}&=\sum_{i,j,k}\int_{x_{i,\rm min}}^1 \frac{dx_i}{x_i}f_{i/P}(x_i,\mu)\int_{x_{j,\rm min}}^1 \frac{dx_j}{x_j}f_{j/P}(x_j,\mu) \int_{z_{\text{min}}}^1 \frac{dz}{z}\, \mathcal{H}_{ij}^k(x_i,x_j,\pt/z,\eta,\mu)\nn\\
    &\times J_k\left(z, \ln\frac{\pt^2 R^2}{z^2 \mu^2},\mu\right)\,,
\end{align}
where the lower integration bounds on the energy fractions are functions of $\pt$ and $\eta$, where
\begin{align}
x_{i,\min }=1-\frac{1-Z}{V}, \quad x_{j,\min }=\frac{1-V}{1+(1-V-Z) / x_i}, \quad z_{\min }=\frac{1-V}{x_j}-\frac{1-V-Z}{x_i}\,,
\end{align}
and
\begin{align}
V=1-\frac{2 \pt}{\sqrt{s}} e^{-\eta}, \quad Z=\frac{2 \pt}{s} \cosh \eta\,.
\end{align}

In the factorization theorem, each ingredient describes the dynamics at one scale.
For instance, $f_{i/P}$ represents the PDF that gives the probability of finding a parton $i$ inside a proton $P$. The process-dependent hard coefficients $\mathcal{H}_{ij}^k$ characterize the underlying hard-scattering process producing a jet-initiating parton $k$. These coefficients have been numerically computed at NNLO accuracy using the approach outlined in Refs.~\cite{Czakon:2021ohs, Czakon:2022pyz}, yielding fully differential single-inclusive partonic cross sections. Finally, $J_k$ represents the (semi-)inclusive jet function introduced in Ref.~\cite{Kang:2016mcy}. Notably, as discussed in detail in Ref.~\cite{Lee:2024icn}, the convolution structure for the inclusive jet function is modified relative to the usual DGLAP convolution as previously believed. In App.~\ref{sec:power-corrections} and~\ref{sec:non-perturbative}, respectively, we will also discuss the perturbative and non-perturbative power corrections to Eq.~\eqref{eq:factorization}.

The factorization for single-inclusive hadron production takes on a similar form relative to the single-inclusive jet production, but with the usual DGLAP convolutions:
\begin{align}
\label{eq:hadfact}
    \frac{d\sigma_h}{d\pt d\eta}&=\sum_{i,j,k}\int_{x_{i,\rm min}}^1 \frac{dx_i}{x_i}f_{i/P}(x_i,\mu)\int_{x_{j,\rm min}}^1 \frac{dx_j}{x_j}f_{j/P}(x_j,\mu) \int_{z_{\text{min}}}^1 \frac{dz}{z}\, \mathcal{H}_{ij}^k(x_i,x_j,\pt/z,\eta,\mu)\nn\\
    &\times D_{k\to h}\left(z,\mu\right)\;,
\end{align}
where $D_{k\to h}(z,\mu)$ denotes the single-inclusive fragmentation function describing hadron production from a fragmenting parton $k$. Consequently, any computational framework capable of evaluating such hadron fragmentation processes can, in principle, be extended to compute fragmentation into small-radius jets with only minor modifications to the convolution procedure. In this study, we use the C++ implementation of the \textsc{Stripper} framework, which provides a fully general subtraction scheme for fixed-order cross section computations through NNLO QCD. This framework allows us to calculate both the inclusive jet cross sections at fixed-order precision and the resummed contributions given in Eq.~\eqref{eq:factorization}. For inclusive jet production, tree-level matrix elements with up to six partons are required and taken from the \textsc{AvH} library~\cite{Bury:2015dla}. The necessary one-loop amplitudes with up to five partons are taken from the \textsc{OpenLoops2} library~\cite{Buccioni:2019sur}. The four-parton two-loop matrix elements have been taken from Ref.~\cite{Broggio:2014hoa}. The framework has been expanded to allow for fragmentation processes into hadrons in Refs.~\cite{Czakon:2021ohs, Czakon:2022pyz,Czakon:2024tjr, Czakon:2025yti}, and additional modifications needed for this work are described in more detail in Sec.~\ref{subsec:stripper}.

\subsection{Fragmentation formalism in \textsc{Stripper} }\label{subsec:stripper}

In this subsection, we elaborate on the calculation of hard function and its convolution with the inclusive jet function within the \textsc{Stripper} framework. Since the LP factorized form of the production cross section of small-radius jets given in Eq.~\eqref{eq:factorization} is of the similar form as the factorization formula for hadron fragmentation process given in Eq.~\eqref{eq:hadfact}, the implementation of fragmentation in \textsc{Stripper} can be extended to compute factorized contribution for jet spectra as well. In essence, this can be achieved by choosing a set of `fragmentation functions' equal to the inclusive jet functions and treating the produced `hadron' as a jet.

There are two subtleties to this. The first is that, unlike fragmentation functions, the inclusive jet functions depend on the kinematics of the initiating parton. For example, heavy-quark fragmentation functions depend only on the momentum fraction, the heavy-quark mass and the factorization scale. None of these are explicitly sensitive to the kinematics of the fragmenting parton or indeed any aspect of the hard scattering process. Of course, the factorization scale can be chosen to be a dynamical scale that depends on some aspects of the full event, and the implementation in \textsc{Stripper} fully supports any choice of scale. The inclusive jet functions, on the other hand, depend on the momentum fraction, the factorization scale, the jet radius and the $\pt$ of the final jet. This last point is what differentiates the present application from the fragmentation calculations considered in previous works. Obviously, the $\pt$ of the final jet is the product of the momentum fraction and the $\pt$ of the parton initiating the jet, meaning the inclusive jet functions contain an explicit dependence on the kinematics of the hard scattering process. In a sense, $\pt R/z $ plays the same role for the inclusive jet functions as the heavy-quark mass does for the heavy-quark fragmentation functions. However, while the latter is a parameter which is fixed throughout the computation, the former changes from event to event. Previously, the implementation of fragmentation in \textsc{Stripper} did not allow for such `event-dependent parameters', so the implementation has been extended to support this.

The second subtlety relates to the analytic form of the inclusive jet functions. The initial implementation of fragmentation in \textsc{Stripper} presented in Ref.~\cite{Czakon:2021ohs} assumed that the fragmentation functions are regular functions of $x$ on $x\in(0,1]$. However, the inclusive jet functions are distribution-valued, containing $\delta$-distributions and plus-distributions. The implementation in \textsc{Stripper} was already extended to support convolutions with such distribution-valued functions in Ref.~\cite{Czakon:2024tjr}. Conceptually, numerical convolutions with distribution-valued functions are not new to the phenomenology community, nor are they particularly complicated. However, since this case differs from the standard hadron fragmentation scenario, we will outline its implementation here.

Let us first consider convolutions with regular functions, as is commonly needed for calculations involving fragmentation. At the partonic level, an MC event is determined by its weight and its kinematics. Focusing on just the kinematics of the fragmenting particle, we can thus schematically write the $i$th parton-level event as $\{w_i,p_i\}$. In general, a single event will consist of multiple contributions of different weight and kinematics, c.f.~subtraction terms in the context of subtraction schemes. However, this does not affect the discussion. The hadron-level event would then be $\{w_iD(x_i),x_ip_i\}$, where we multiplied the parton's momentum with the momentum fraction $x_i$ to obtain the hadron's momentum and multiplied the weight by the relevant fragmentation function evaluated at $x_i$. $x_i$ can simply be randomly sampled from $(0,1]$. This simple modification is all it takes to perform convolutions with regular functions. A convolution with $\delta(1-x)$ is trivial: it does not change the event at all, i.e.~the hadron-level event would still be $\{w_i,p_i\}$. A convolution with a plus-distribution, e.g.~$(1-x)^{-1}_{{}+{}}$, gives two contributions: $\{w_i/(1-x_i),x_ip_i\}$, corresponding to the bulk of the plus-distribution, and $\{-w_i/(1-x_i),p_i\}$, corresponding to the endpoint, or subtraction term, of the plus-distribution. Note that attempting to integrate either of these terms over $x_i$ individually leads to a non-integrable singularity. But when integrated together, the singular behaviours as $x\to1$ cancel, since the kinematics of the two contributions become identical in that limit, meaning the two divergent contributions necessarily land in the same bin of any histogram. This is in complete analogy to the way subtraction schemes for the integration of real radiation contributions in higher-order calculations work.

So, concretely: if a partonic cross section is to be convolved with a function $f(x)+g\:\delta(1-x)+h_{{}+{}}(x)$, where $f(x)$ is regular, and the $i$th partonic event, generated as usual, is represented by $\{w_i,p_i\}$, then the convolved cross section is obtained by randomly drawing $x_i\in(0,1]$ and summing the contributions $\{w_if(x_i),x_ip_i\}$, $\{w_ig,p_i\}$, $\{w_ih(x_i),x_ip_i\}$ and $\{-w_ih(x_i),p_i\}$ for each event. If a partonic event consists of multiple contributions, then this process is simply repeated for every partonic contribution, using the same value of $x_i$ throughout a given partonic event.

On top of the two subtleties described above, one of course needs to actually implement the expressions for the inclusive jet functions in the code through the desired orders. Now that all of this has been done for \textsc{Stripper}, it is fully capable of computing any jet cross section at NNLO+NNLL, provided it could previously compute it at NNLO. In practice, this means every process for which the two-loop amplitudes are known. Note that the implementation described here is fully general, and the implementation of convolutions with other, possibly more general functions, such as those needed to compute energy correlators, would be reasonably straightforward.

\subsection{Inclusive jet function}
The inclusive jet function $J_i\left(z,\ln\frac{\pt^2 R^2}{z^2\mu^2},\mu\right)$ in the factorization given in Eq.~\eqref{eq:factorization} describes the probability of producing a final-state jet with energy fraction $z$.
Recently, it was shown~\cite{Lee:2024icn,Lee:2024tzc} that the inclusive jet function obeys the following RG evolution equation given as
\begin{align}
\label{eq:jetRG_z}
    \frac{d \vec{J}\left(z,\ln\frac{\pt^2 R^2}{z^2\mu^2},\mu \right)}{d \ln \mu^2} = \int_z^1 \frac{dy}{y}  \vec{J}\left(\frac{z}{y},\ln\frac{y^2\pt^2 R^2}{z^2\mu^2},\mu \right) \cdot \widehat P_T(y)\,,
\end{align}
where $\widehat P_T$ is the singlet timelike splitting kernel matrix, known up to three-loop accuracy~\cite{Mitov:2006ic,Mitov:2006wy,Moch:2007tx,Chen:2020uvt}. This result contrasts with earlier proposals in the literature~\cite{Kang:2016mcy,Dai:2016hzf}, which suggested that the inclusive jet function evolves according to the standard DGLAP equation familiar from single-inclusive hadron production. A notable difference from the standard DGLAP evolution is the appearance of the convolution variable $y$ in the logarithmic argument on the right-hand side of Eq.~\eqref{eq:jetRG_z}. (This deviation from the standard DGLAP structure has also been observed in~\cite{vanBeekveld:2024jnx,vanBeekveld:2024qxs}.) Taking the $N$-th moment, $\vec{J} \left(N,\ln\frac{\pt^2 R^2}{z^2\mu^2},\mu\right) \equiv \int_0^1 dz\, z^{N}\,\vec{J} \left(z,\ln\frac{\pt^2 R^2}{z^2\mu^2},\mu\right)$, it is noteworthy that, in moment space, the RG evolution equation for the inclusive jet function shares structural similarities with the collinear limit of the projected $N$-point energy correlator~\cite{Dixon:2019uzg,Chen:2020vvp}:
\begin{align}
  \label{eq:incljetRG}
\frac{d \vec{J}\left(N,\ln\frac{\pt^2R^2}{z^2\mu^2}, \mu\right) }{d \ln \mu^2} = \int_0^1 dy\, y^{N} \vec{J} \left(N,\ln\frac{y^2 \pt^2R^2}{z^2\mu^2}, \mu\right) \cdot \widehat P_T(y) \,.
\end{align}

In order to perform the collinear resummation, we need to solve the jet RGE in Eq.~\eqref{eq:jetRG_z}. We follow Ref.~\cite{Lee:2024tzc} and solve the inclusive-jet RG evolution iteratively to NNLL accuracy. Explicitly, we use the following ansatz for the renormalized jet function,
\begin{align}
\label{eq:Jren}
J_i\left(z,\ln\frac{\pt^2 R^2}{z^2\mu^2},\mu \right) =& \sum_{n=0}^\infty \sum_{m=0}^n \frac{a_s^n\, L_z^m}{m!}J_i^{(n,m)} = \delta(1-z) + a_s \left[J_i^{(1,0)} + \textcolor{blue}{J_i^{(1,1)}}L_z\right] \nn\\
&\hspace{-3cm}+ a_s^2 \left[J_i^{(2,0)} + \textcolor{blue}{J_i^{(2,1)}}L_z + \textcolor{blue}{J_i^{(2,2)}}\frac{L_z^2}{2}\right]\nn\\
&\hspace{-3cm}+a_s^3 \left[J_{i}^{(3,0)}+\textcolor{blue}{J_{i}^{(3,1)}}L_z + \textcolor{blue}{J_{i}^{(3,2)}}\frac{L_z^2}{2}+ \textcolor{blue}{J_{i}^{(3,3)}}\frac{L_z^3}{3!}\right] \nn\\
&\hspace{-3cm}+a_s^4 \left[J_{i}^{(4,0)}+\textcolor{blue}{J_{i}^{(4,1)}}L_z + \textcolor{blue}{J_{i}^{(4,2)}}\frac{L_z^2}{2}+ \textcolor{blue}{J_{i}^{(4,3)}}\frac{L_z^3}{3!}+ \textcolor{blue}{J_{i}^{(4,4)}}\frac{L_z^4}{4!}\right] \nn\\
&\hspace{-3cm}+a_s^5 \left[J_{i}^{(5,0)}+\textcolor{blue}{J_{i}^{(5,1)}}L_z + \textcolor{blue}{J_{i}^{(5,2)}}\frac{L_z^2}{2}+ \textcolor{blue}{J_{i}^{(5,3)}}\frac{L_z^3}{3!}+ \textcolor{blue}{J_{i}^{(5,4)}}\frac{L_z^4}{4!}+ \textcolor{blue}{J_{i}^{(5,5)}}\frac{L_z^5}{5!}\right] + \mathcal{O}(a_s^6)\,,
\end{align}
to $5$-loop order, where $L_z = \ln z^2\mu^2/(\pt^2R^2)$ and $a_s = \alpha_s/(4\pi)$. Note that each $J_i^{(n,m)}$ is a function of momentum fraction $z$ only. It is straightforward to generalize this to higher-loop orders and the terms in blue are predicted by RG evolution. Regarding log counting, the N$^k$LL accuracy refers to the terms $J_i^{(n,m)}$ with $n\geq m \geq n-k$.

We then obtain the iterative solution up to NNLL given as
\begin{align}
\label{eq:NNLLiter}
J_i^{(m,m)}(z) &= P_{ki}^{(0)} \otimes J_k^{(m-1,m-1)} + (m-1)\beta_0 J_i^{(m-1,m-1)}\,, \hspace{2cm}  \text{ for } m\geq 1\,,\nn\\
J_i^{(m,m-1)}(z) &= P_{ki}^{(1)}\otimes J_k^{(m-2,m-2)}  + P_{ki}^{(0)}\otimes J_k^{(m-1,m-2)}+ (m-1)\beta_0 J_{i}^{(m-1,m-2)} \nn\\
&+ (m-2)\beta_1 J_{i}^{(m-2,m-2)}- 2P_{ki}^{(0)}\otimes (J_{k}^{(m-1,m-1)}\textcolor{red}{\ln y})\,, \hspace{0.85cm}\text{ for } m\geq 2\,,   \nn\\
J_i^{(m,m-2)}(z) &= P_{ki}^{(2)}\otimes J_i^{(m-3,m-3)}+P_{ki}^{(1)}\otimes J_i^{(m-2,m-3)}+P_{ki}^{(0)}\otimes J_i^{(m-1,m-3)}\nn\\
&+(m-1)\beta_0 J_i^{(m-1,m-3)}+(m-2)\beta_1 J_i^{(m-2,m-3)}+(m-3)\beta_2 J_i^{(m-3,m-3)} \nn\\
&-2P_{ki}^{(1)}\otimes (J_i^{(m-2,m-2)} {\color{red}\ln y})-2 P_{ki}^{(0)}\otimes (J_i^{(m-1,m-2)}{\color{red}\ln y})\\
& + 4 P_{ki}^{(0)}\otimes (J_i^{(m-1,m-1)} {\color{red}\ln^2 y}),\, \hspace{4.9cm}\text{ for } m\geq 3 \,,\nn
\end{align}
where we also expand the splitting kernels and the beta function as
\begin{equation}
    P_{ij}=\sum_{n=0} a_s^{n+1} P_{ij}^{(n)},\, \quad \frac{da_s(\mu)}{d\ln\mu}=\frac{\beta(a_s(\mu))}{4\pi}=-2 a_s(\mu) \left[a_s(\mu)\beta_0+a_s(\mu)^2 \beta_1+\cdots \right]\,.
\end{equation}
Here, the short notation for DGLAP convolution $\otimes$ is defined as
\begin{align}
    [A\otimes B](z) &\equiv \int_z^1 \frac{dy}{y} A(y) B(z/y) = \int_z^1 \frac{dy}{y} A(z/y) B(y)\,,
\end{align}
The presence of $\textcolor{red}{\ln y}$ terms in the solution arises from the difference between the inclusive jet RG evolution in Eq.~\eqref{eq:jetRG_z} from the standard DGLAP structure found in the literature.

With the blue terms fixed by the RG evolution, we now need the fixed order parts given by $J_i^{(1,0)}(z)$ and $J_i^{(2,0)}(z)$ to achieve full NNLL accuracy. The NLO jet functions $J_i^{(1,0)}(z)$ are known analytically~\cite{Kang:2016mcy} and are given as
\begin{align}
J_q^{(1,0)}(z)= & -2 \ln z\left[P_{q q}^{(0)}(z)+P_{g q}^{(0)}(z)\right] \\
& -\left\{2 C_F\left[2\left(1+z^2\right)\left(\frac{\ln (1-z)}{1-z}\right)_{+}+(1-z)\right]\right.\nn \\
& \left.-\delta(1-z) 2 C_F\left(\frac{13}{2}-\frac{2 \pi^2}{3}\right)+2 P_{g q}^{(0)}(z) \ln (1-z)+2 C_F z\right\}\,,\nonumber\\
J_g^{(1,0)}(z)= & -2 \ln z\left[P_{g g}^{(0)}(z)+2 N_f P_{q g}^{(0)}(z)\right] \nn\\
& -\left\{\frac{8 C_A\left(1-z+z^2\right)^2}{z}\left(\frac{\ln (1-z)}{1-z}\right)_{+}-2 \delta(1-z)\left(C_A\left(\frac{67}{9}-\frac{2 \pi^2}{3}\right)\right.\right.\nn \\
& \left.\left.-T_F N_f\left(\frac{23}{9}\right)\right)+4 N_f\left(P_{q g}^{(0)}(z) \ln (1-z)+2 T_F z(1-z)\right)\right\}\,.\nn
\end{align}
Explicit full two-loop jet function constants have not been computed thus far, whereas their pole terms were computed for the first time recently in Ref.~\cite{Lee:2024icn}. However, as we will describe below, with the full fixed-order NNLO computation of the small-$R$ radius jet process available, we can extract the two-loop jet function constants by considering the small-$R$ limit and comparing it with the resummation calculation, where only the two-loop jet constants remain unknown. To test our extraction, we will compare our extractions with the NLL threshold expansions of the two-loop jet functions. In the threshold limit $z\to 1$, or equivalently, the large-$N$ limit in moment space, the moments of the two-loop constants, $J_i^{(2,0)}(N) = \int_0^1 dz\,z^N J_i^{(2,0)}(z)$, are given as
\begin{align}
\label{eq:jetthreshold}
J_q^{(2,0)} (N)|_{N\gg 1}=& C_A C_F \left(\frac{-11}{3}\pi^2 \ln\bar{N} + \frac{-268+12\pi^2}{9}\ln^2\bar{N} - \frac{88}{9}\ln^3\bar{N}\right) \nonumber\\
&+ C_Fn_FT_F \left(\frac{4}{3}\pi^2 \ln \bar{N} +\frac{80}{9}\ln^2 \bar{N} +\frac{32}{9}\ln^3 \bar{N}\right) \nonumber\\
&+ C_F^2\left(\frac{1}{4}\pi^2(-26+3\pi^2) +(-52+8\pi^2)\ln^2\bar{N} + 8\ln^4\bar{N}\right)\,,\\
J_g^{(2,0)}(N)|_{N\gg 1}=& C_A^2 \left(\frac{\pi^2}{36}(-268+27\pi^2) - \frac{11}{3}\pi^2\ln\bar{N} + \frac{4}{3}\left(-67+7\pi^2\right)\ln^2\bar{N} \right.\nonumber\\
&\left.\quad\quad-\frac{88}{9}\ln^3\bar{N} + 8\ln^4\bar{N}\right) \nonumber\\
&+ C_An_FT_F \left(\frac{23\pi^2}{9}+\frac{4}{3}\pi^2\ln\bar{N} +\frac{88}{3}\ln^2\bar{N}+\frac{32}{9}\ln^3\bar{N}\right)\,,
\end{align}
where $\bar{N} = e^{\gamma_E}N$.

\subsubsection{Numerical extraction of the NNLO inclusive jet function}

As explained above, the inclusive jet functions are currently only know analytically through NLO. However, in order to perform NNLL resummation, the NNLO corrections are needed as well. The terms enhanced by powers of $\ln R$ can be obtained from the lower-order contributions through DGLAP evolution. Similarly, terms proportional to powers of $\ln(\mu_J/\mu_R)$ can be obtained from lower orders as well. This means only the terms independent of $R$ and $\mu_J$ are unknown. The sum of those terms will be referred to as the `NNLO jet constants' $J_i^{(2,0)}$ given in Eq.~\eqref{eq:NNLLiter}. The NNLO jet constants can then be obtained numerically by comparing the fixed order predictions with the factorized contribution in Eq.~\eqref{eq:factorization}:
\begin{align}
\sum_{i}d\sigma_i^\text{LO}\otimes a_s^2J_i^{(2,0)} =&\bigg[d\sigma^\text{NNLO} -\sum_{i}d\sigma_i^\text{NNLO}\nn\\
&\hspace{-1cm}-\sum_{i}d\sigma_i^\text{LO}\otimes  a_s^2\left(J_i^{(2,1)}(L+\ln z^2)+J_i^{(2,2)}\frac{(L+\ln z^2)^2}{2}\right) \nn\\
    &\hspace{-1cm}- \sum_{i}d\sigma_i^\text{NLO}\otimes a_s\left(J_i^{(1,0)}+J_i^{(1,1)}(L+\ln z^2)\right)\bigg](1+\mathcal{O}(R^2))\;,
\end{align}
where $L = \ln \mu^2/(\pt^2R^2)$ and $\otimes$ indicates the usual DGLAP convolution notation. We emphasize again that extra $\ln z^2$ terms arise from the modified convolution relative to the usual DGLAP convolution structure. Here, we used shorthand notation $d\sigma_i$ to refer to the combination of the RHS of Eq.~\eqref{eq:factorization} that convolves with the jet function. On the other hand, the notation $d\sigma^{\rm NNLO}$ indicates the NNLO fixed order computation of the inclusive jet production with desired $\pt$ and $\eta$. 
In the moment space, this becomes
\begin{align}
\label{eq:momspJetconst}
    \sum_{i}d\sigma_i^\text{LO}(N)&a_s^2J_i^{(2,0)}(N) =\bigg[d\sigma^\text{NNLO}(N) - \sum_{i}d\sigma_i^\text{NNLO}(N)\nn \\
    &- \sum_{i}d\sigma_i^\text{LO}(N)  a_s^2\left(J_i^{(2,1)}(N)L+J_i^{(2,2)}(N)\frac{L^2}{2}\right)\nn \\
    &- \sum_{i}2d\dot{\sigma}_i^\text{LO}(N) \left(a_s J_i^{(1,1)}(N)+ a_s^2 J_i^{(2,1)}(N)+a_s^2 J_i^{(2,2)}(N)L\right)\nonumber\\
    &- \sum_{i}2d\ddot{\sigma}_i^\text{LO}(N)  a_s^2J_i^{(2,2)}(N)- \sum_{i}d\sigma_i^\text{NLO}(N)a_sJ_i^{(1,0)}(N)\nonumber\\
    &- \sum_{i}d\sigma_i^\text{NLO}(N) a_sJ_i^{(1,1)}(N)L\bigg](1+\mathcal{O}(R^2))\;,
\end{align}
where
\begin{align}
d\dot{\sigma}_i(N) =& \int_0^1 dz\,z^N \ln z \,d\sigma_i(z)\,,\nn\\
d\ddot{\sigma}_i(N) =& \int_0^1 dz\,z^N \ln^2 z \,d\sigma_i(z)\,,
\end{align}
are logarithmic moments of hard coefficients that appear from modified convolution structure.

All terms on the RHS of Eq.~\eqref{eq:momspJetconst} can be computed numerically. If there were only one jet function, e.g.~only $J_g^{(2,0)}$, then $J_{g}^{(2,0)}(N)$ could be obtained by simply dividing both sides by $d\sigma_{g}^{\rm LO}(N)$. In practice, both $J_g^{(2,0)}(N)$ and $J_q^{(2,0)}(N)$ are non-zero, so they cannot be obtained directly. Instead, we consider multiple cross sections, constructed to resolve this degeneracy between gluon- and quark-initiated jets, and perform a fit to obtain the values of $J_{g}^{(2,0)}(N)$ and $J_{q}^{(2,0)}(N)$ for a range of values of $N$ that covers what is needed for phenomenology. Note that since we rely on the factorization of convolutions into simple products after taking the moments, care must be taken not to break this factorization. This forces us to use factorization and renormalization scales and phase-space cuts which are independent of the jet kinematics. Here, we choose the central scales to be 25 GeV, including also the usual variations by factors of two to verify that the final result does not depend on the choice of scale.

We implement two ways to lift degeneracy between the quark and gluon two-loop jet constants.  First, we compute double-differential moments in (non-absolute) rapidity and $\hat{H}_T$, the scalar sum of partonic transverse momenta. Note that the latter does not break the factorization in moment space, since cutting on $\hat{H}_T$ modifies the partonic cross section only. This is in contrast to binning in the jet $\pt$, which would break the factorization. Specifically, we put the rapidity bin edges at $\{-\infty,\:-8,\:-6,\:-4,\:-2,$ $\:0,\:2,\:4,\:6,\:8,\:+\infty\}$ and the $\hat{H}_T$ bin edges at $\{25,\:50,\:100,\:200,\:400,$ $\:800,\:1600,\:3200,\:6500\}$ GeV. Note that $\hat{H}_T$ cannot exceed 6500 GeV as we perform our computation for the 13 TeV LHC. Second, we split up the cross section according to the initial state channels, i.e.~we act as if we can turn on and off specific PDFs for each proton.  While turning PDFs on or off does not correspond to physical cross sections, this provides a convenient handle for accurate numerical extraction. Specifically, we consider four channels: $gg$, $gq$ (proton 1 only contains gluons, proton 2 contains all (anti-)quarks), $\overline{q}q$ (proton 1 only contains anti-quarks, proton 2 only contains quarks, the flavours of the colliding (anti-)quarks must be the same) and finally the sum over all channels involving only (anti-)quarks, excluding $q\overline{q}$ and $\overline{q}q$ (i.e.~there can be no gluons in the final state at LO, meaning this contribution does not contain terms proportional to the gluon jet function constant at NNLO). Combining the double-differential moments with the split channels allows us to essentially fully disentangle the quark and gluon jet function constants, with the final uncertainties on the extracted moments having correlations smaller than 0.25.

\begin{figure}[h]
	\centering
	\includegraphics[width=0.49\textwidth]{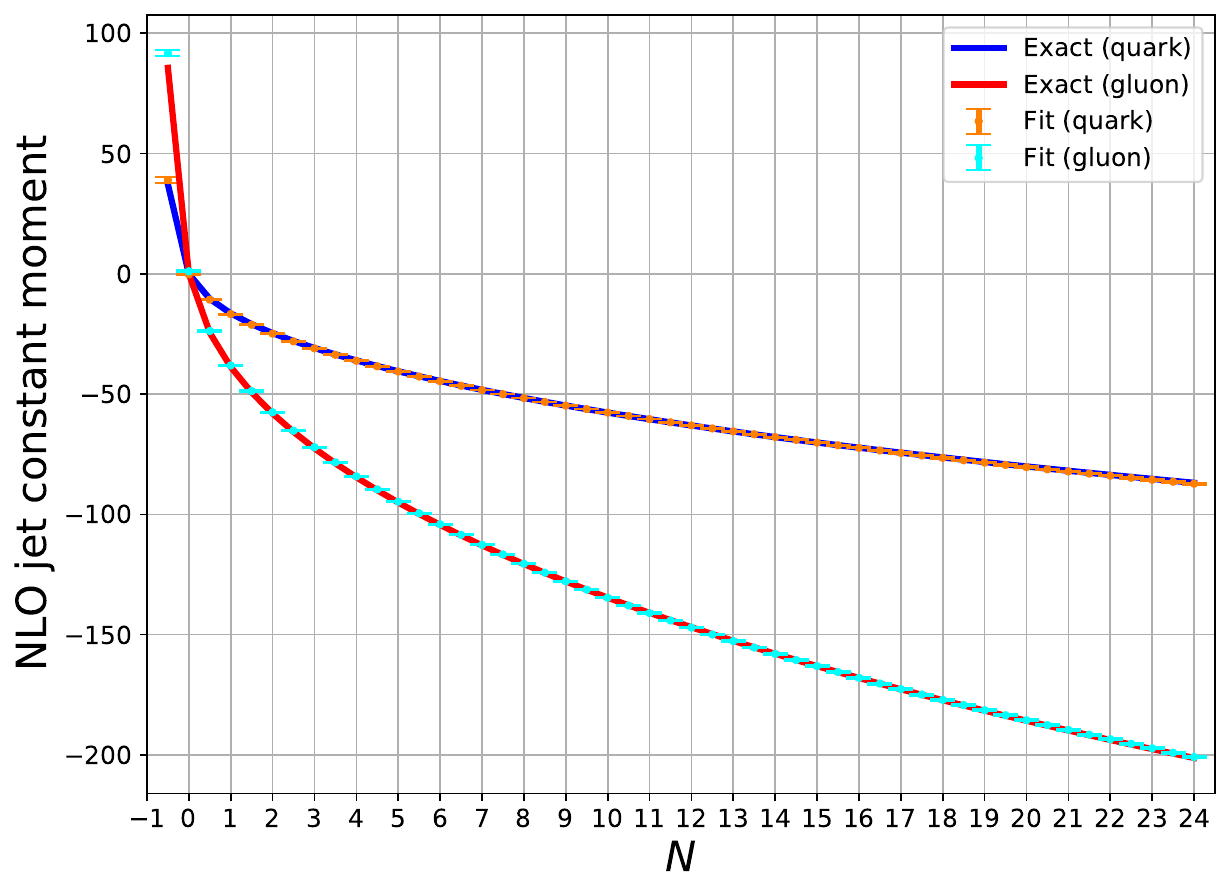}
	\includegraphics[width=0.49\textwidth]{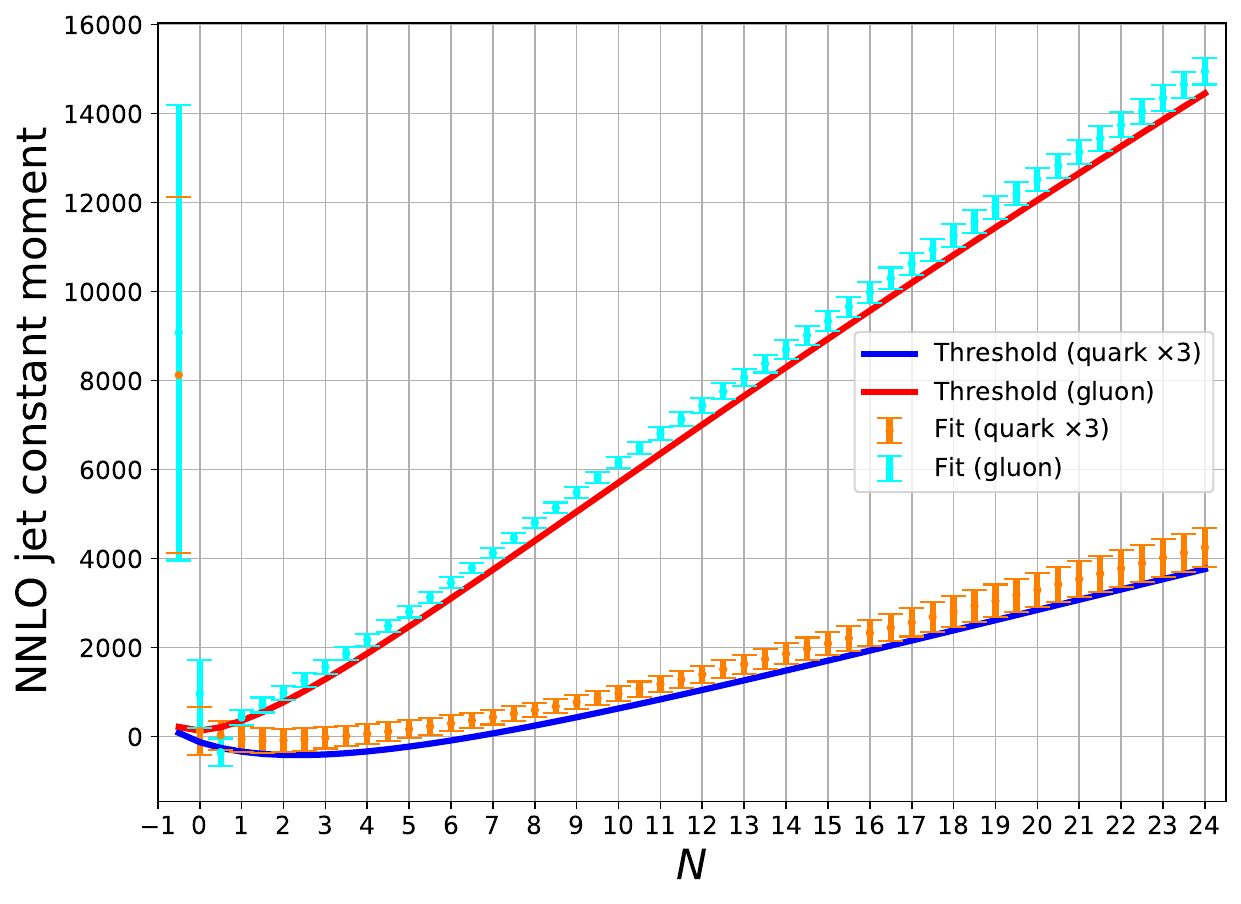}
	\caption{Numerically extracted moments of the NLO (left) and NNLO (right) jet function constants, together with the analytic moments of the known exact result (at NLO) or the NLL threshold expansion (at NNLO). The error bars indicate the fit uncertainty. Note that the NNLO quark jet function constant has been multiplied by 3 for clarity.}
	\label{fig:MomentFit}
\end{figure}

To obtain the results, we now simply perform a standard $\chi^2$ minimization for each value of $N$, where $\chi^2$ is defined as usual, summing over all bins of all four double-differential distributions and assuming that the MC errors for different $y$ and $\hat{H}_T$ bins are uncorrelated. The central values correspond to the position of the minimum of $\chi^2$, while the fit uncertainties (and their correlations) are obtained from the inverse Hessian. The results of this fit are shown in Fig.~\ref{fig:MomentFit}. The left panel shows the result of applying this procedure to extract the NLO jet constant. Clearly, the moments obtained from the fit line up perfectly with the exact result. The right panel shows the NNLO fit results and compares them to the approximate NNLO jet constant derived from the threshold expansion, as given in Eq.~\eqref{eq:jetthreshold}. Deviations of 5-10\% are observed, consistent with both the size of the fit uncertainties and the expected size of the terms missing from the threshold expansion. To minimize the impact of power corrections in $R$, these result were obtained using $R=0.1$. It was checked that repeating the fit for $R=0.2$ leads to a shift in the extracted moments smaller than the fit uncertainties.

The fit uncertainty could in principle be further reduced by running the MC integration for longer. However, it was found that the total contribution stemming from the NNLO jet constants is at most 3\% of the NNLO+NNLL cross sections presented here, and typically considerably less than that. As a result, the uncertainties in the final results due to these fit uncertainties are entirely negligible compared to other sources of uncertainty. The numerical results of the fit are included as an ancillary file with the e-print version of this paper.

\subsubsection{Using moments to get transverse momentum spectra}

The moments of the NNLO jet function constants have now been extracted, but these are not what one traditionally needs to obtain differential cross sections. If one is interested in e.g.~computing the inclusive jet $\pt$ spectrum, one needs to compute the convolution of the partonic cross sections and the inclusive jet functions. However, this requires knowledge of the jet functions in $x$-space, not $N$-space. If one knew all moments analytically, one could simply perform a Mellin inversion. But since only a finite number of moments have been extracted with only finite precision, this is not possible here. Instead, we rely on the following approximation.

It was pointed out a long time ago \cite{Frixione:1997ma,Nason:1999ta} that when computing a convolution involving a steeply falling spectrum, as is the case for jet $\pt$ spectra, only a single moment of the other function tends to contribute:
\begin{align}
\label{eq:convolsteep}
    \bigg(\frac{d\sigma}{d\pt}\otimes J\bigg)(\pt) &{}= \int_0^1\frac{1}{x}\frac{d\sigma}{d\pt}(\pt/x)J(x) dx = \int_0^1\frac{C}{x}\bigg(\frac{x}{\pt}\bigg)^n J(x) dx\notag\\&{}= \frac{C}{\pt^n}\int_0^1x^{n-1} J(x) dx = \frac{CJ(n-1)}{\pt^n} = J(n-1)\frac{d\sigma}{d\pt}(\pt)\;,
\end{align}
where we assumed that the partonic cross section behaves like $1/\pt^n$ for some constant $n$. In other words: if the partonic $\pt$ spectrum behaves like a power law w.r.t.~$\pt$ with exponent $n$, then the convolution with the jet function can be obtained by simply multiplying by its `$n-1$'th moment. (Note that the contribution associated with the two-loop jet function constants are in the simple DGLAP convolution form as Eq.~\eqref{eq:convolsteep}).

Of course, the real partonic cross sections do not show such perfect power law behaviour. In general, the exponent $n$ will also depend on $\pt$. However, if $n$ varies sufficiently slowly with $\pt$, then one can use, to very good approximation, the following formula:
\begin{equation}
    \bigg(\frac{d\sigma}{d\pt}\otimes J\bigg)(\pt) \approx J(n(p_T)-1)\frac{d\sigma}{d\pt}(\pt)\;,
\end{equation}
where $n$ now depends on $\pt$.

To see why this is a good approximation consider the following. Assume the convolution is to be computed for the point $\pt=p_{T,0}$. The partonic cross section will deviate from a power law with $n=n(p_{T,0})$ as the transverse moment at which it is evaluated is either increased or decreased. However, partonic transverse momenta below $\pt=p_{T,0}$ do not contribute, since the momentum fraction is restricted to $(0,1]$. Higher momenta do contribute, but, by assumption, the dependence of $n$ on $\pt$ is weak. Therefore, a significant deviation from $n=n(p_{T,0})$ is only observed for partonic $\pt\gg p_{T,0}$. Since it was also assumed that the spectrum is steeply falling (indeed typically $n\gtrsim5$), larger values of $\pt$ will yield increasingly smaller contributions to the final result. That is, by the time $n$ has deviated significantly from $n(p_{T,0})$, the cross section has been suppressed so much by the large value of $\pt$ that its contribution is negligible anyway.

Within our resummation scheme, we only need the convolution of the NNLO jet function constants with the LO partonic cross sections; the latter can be obtained trivially. We then estimate values of $n(\pt)$ for each partonic cross section by taking ratios of consecutive $\pt$ bins and constructing a simple linear interpolating function. Note that care must be taken to divide out any factors of $\alpha_s$ first when estimating $n$, since $\alpha_s$ should be a multiplicative factor included after the convolution, not before. For the specific double-differential cross sections presented here, $n$ varies between about $4$ and $14$ for the quark cross section and between about $4$ and $20$ for the gluon cross section. All relevant moments can thus be obtained via simple linear interpolation of the moments extracted above. By comparing the true convolution with the NLO jet constants with the corresponding approximation described here, it was found that the approximation is accurate to about 1\%. Again, since the contribution from NNLO jet constants we are approximating here corresponds to at most 3\% of the final result, the resulting error is completely negligible compared to the other sources of uncertainty.

\section{Phenomenology}\label{sec:phenomenology}

In this section, we discuss the phenomenology of jet production as a function of the jet-radius $R$, the transverse momentum $\pt$ and the absolute rapidity $|\y|$ through NNLO+NNLL. The fixed-order and resummation are matched such that only the strictly necessary convolutions to achieve LO+LL, NLO+NLL, and NNLO+NNLL accuracy are computed.

The phase space and histogram binning are taken from the CMS measurement in Ref.~\cite{CMS:2020caw} to facilitate the comparison to data in Sec.~\ref{subsec:cms}. The cross sections considered cover the range $84\textrm{ GeV}<\pt<1588\textrm{ GeV}$ in the transverse momentum, and $\lvert \y\rvert < 2.0$ in the rapidity of the jets. The jet-radii considered are $R=\{0.1,0.2,...,1.2\}$. We used a 15-point scale variation around the central scale choice $\mu_R = \mu_F = \mu_J = \pt$ by a factor of 2 to give estimates of missing higher order uncertainties (MHOU), which are shown as colored bands around the central predictions. The choice of the central scale is motivated by the good perturbative convergence in fixed-order computations of inclusive jet spectra, as discussed in Ref.~\cite{Currie:2018xkj}. For this scale choice, each jet is binned with its own weight according to its transverse momentum within a fixed-order computation. For the resummed prediction, only one ``fragmented'' jet per event needs to be binned.
For the PDFs and $\alpha_s$ we are using the NNLO NNPDF3.1 set~\cite{NNPDF:2017mvq} as implemented in LHAPDF \cite{Buckley:2014ana}.

\begin{figure}[t]
    \centering
    \includegraphics[width=0.5\linewidth]{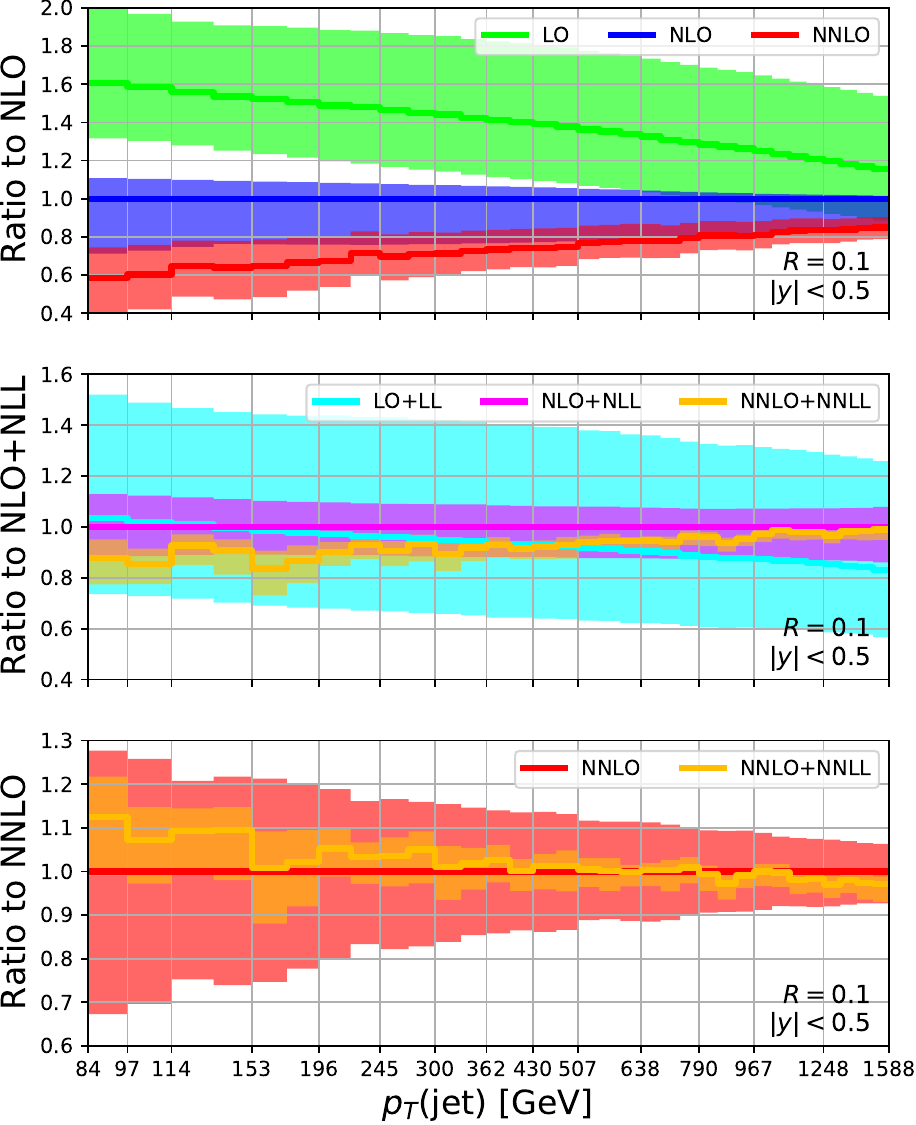}%
    \includegraphics[width=0.5\linewidth]{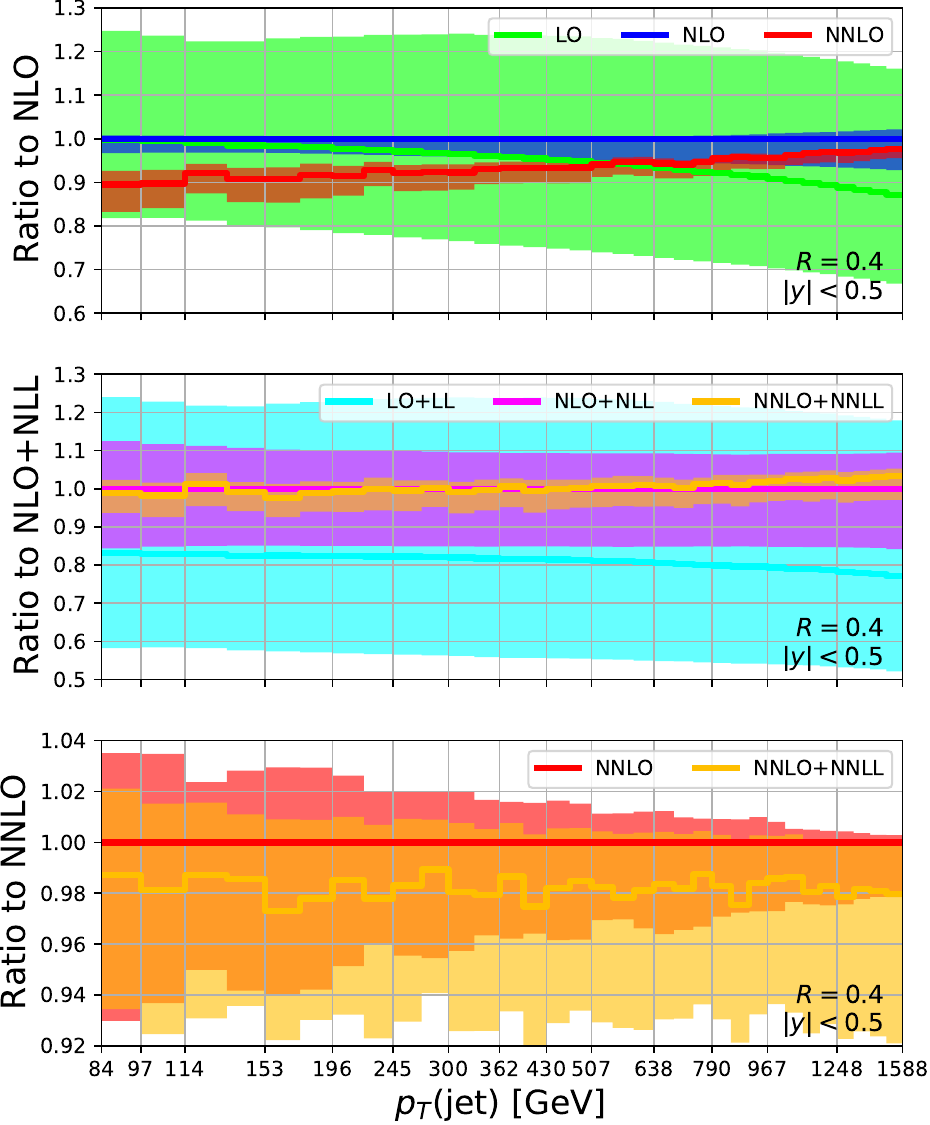}
    \caption{Perturbative predictions for the inclusive jet transverse momentum spectrum for $|\y| < 0.5$. The left and right sides show the results for a jet radius $R=0.1$ and $R=0.4$, respectively. The top panels show fixed-order perturbative predictions (LO - green, NLO - blue, NNLO - red) with respect to NLO QCD. The middle panels compare perturbative predictions matched to small radius resummation (LO+LL - turquoise, NLO+NLL - magenta, NNLO+NNLL orange). The bottom panels show a direct comparison between NNLO QCD and NNLO+NNLL. All bands indicate the envelope obtained from scale variations described in the text.}
    \label{fig:abs-pQCD-1}
\end{figure}
 
\subsection{Perturbative corrections}

\begin{figure}
    \centering
    \includegraphics[width=0.5\linewidth]{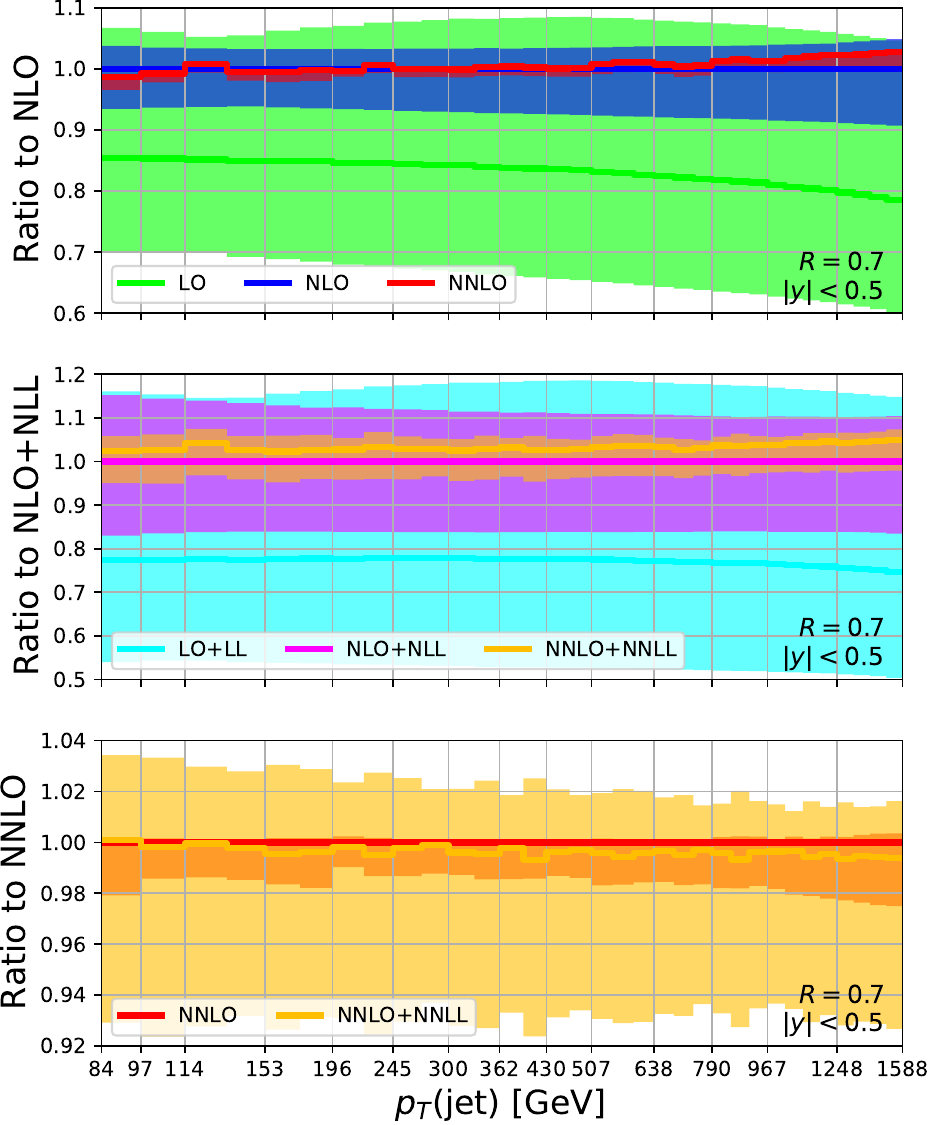}%
    \includegraphics[width=0.5\linewidth]{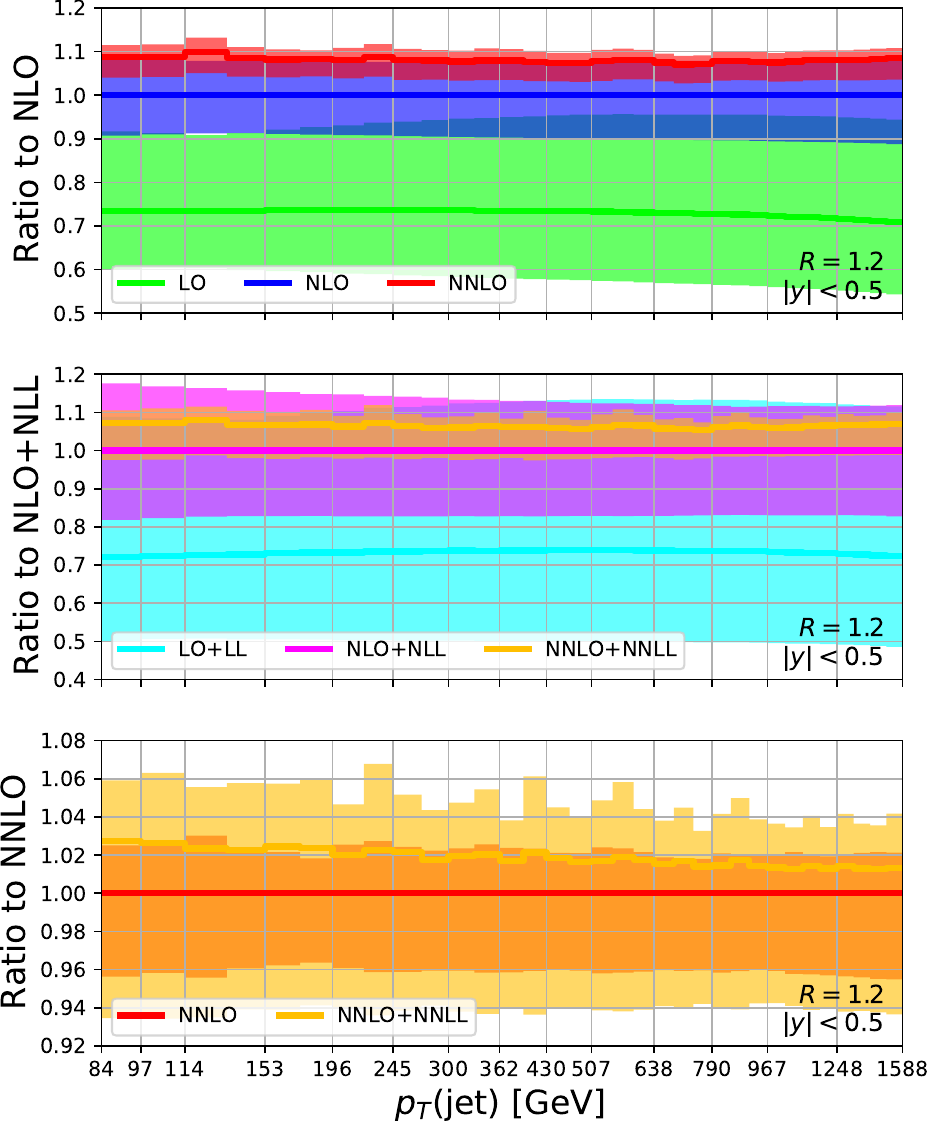}
    \caption{Same as Fig.~\ref{fig:abs-pQCD-1} but for $R=0.7$ and $R=1.2$.}
    \label{fig:abs-pQCD-2}
\end{figure}

First, we discuss the higher-order QCD corrections to the absolute $\pt$ spectrum for different jet radii. In Fig.~\ref{fig:abs-pQCD-1} and Fig.~\ref{fig:abs-pQCD-2} we compare different fixed-order and resummed predictions for jets with radius $R=0.1,0.4,0.7,$ and $1.2$ in a central rapidity region ($|\y|<0.5$)\footnote{The results for all radii and rapidity bins are available in the ancillary files attached to this article.}.

Focusing on fixed-order perturbative predictions (the top panels in Fig.~\ref{fig:abs-pQCD-1} and Fig.~\ref{fig:abs-pQCD-2}), we observe the already known strong dependence of the higher-order corrections on the jet radius $R$ \cite{Currie:2018xkj}. For the smallest radius ($R=0.1$), the NNLO QCD corrections are significant and negative, ranging from $-40\%$ at small to $-20\%$ at high transverse momentum with respect to NLO QCD. The MHOU estimates are about $10-30\%$ at NLO QCD and are slightly reduced to $5-20\%$ at NNLO QCD. The NLO and NNLO QCD predictions barely overlap within their respective uncertainties, indicating slow perturbative convergence. The NNLO QCD corrections are more minor for $R=0.4$, about $-10\%$ at small, and about $-2\%$ at high transverse momentum with respect to NLO QCD. However, the corrections are visibly larger than the MHOU estimates from NLO QCD, which is known to be artificially small. For $R=0.7$, the NNLO QCD corrections are small $\sim \pm 2\%$. Furthermore, the remaining scale dependence is minimal, implying a very small MHOU. For larger radii, we find the corrections generally positive and much flatter in phase space, for example for $R=1.2$ in Fig.~\ref{fig:abs-pQCD-2} we find a $+10\%$ flat correction. In addition, the MHOU estimates stabilize and indicate reasonable perturbative convergence. These findings are entirely consistent with the literature \cite{Currie:2018xkj}.

The situation improves when resummation is included. At small radii, we see stabilization of the perturbative convergence. For $R=0.1$, for example, the second-order corrections lead only to a $-10\%$ change to NLO + NLL, compared to $-40\%$ with fixed-order perturbative QCD. The dependence on the scales is reduced, even though additional variations are included. The second-order corrections also lie within the estimates from NLO+NLL. Similarly, we find a much improved behavior at $R=0.4$. Including the resummation lifts the accidental cancelation of the scale dependence at NLO QCD, now giving a flat and realistic MHOU estimate of $\pm 20\%$. The NNLO+NNLL computation gives only minor corrections with a remaining scale dependence of about $\pm 5\%$. Similar observations hold for larger radii.

In general, we find, visualized in the lower panels of Fig.~\ref{fig:abs-pQCD-1} and Fig.~\ref{fig:abs-pQCD-2}, that the fixed-order perturbative NNLO QCD and the resummed NNLO+NNLL results agree within their respective uncertainty estimates. However, the resummed results have distinguished different scale dependence. As expected, the resummed predictions have a reduced scale dependence for small radii. For intermediate radii, the resummation lifts the artificially small-scale dependence and provides a much more realistic uncertainty estimate of about $\sim 4-5\%$. For large radii, the impact of the resummation is smaller.

In summary, we observe a systematic stabilization of the perturbative series if resummation is included. The resummation lifts various accidental patterns observed in fixed-order perturbative predictions. In particular, it reduces the size of perturbative corrections at small $R$ and solves the issue of the artificial small dependence on the scales at intermediate $R$.

It is interesting to compare to the results of Refs.~\cite{Liu:2017pbb,Liu:2018ktv,Moch:2018hgy}, which presented results for single-inclusive jet cross sections at NLO, jointly resumming jet-radius and threshold logarithms with NLL accuracy. Those works employed a different central scale choice - the leading jet $\pt$ - and so a direct comparison cannot be made. Nevertheless, a qualitative comparison is instructive. Most results can also not be directly compared to simply because of the inclusion of threshold resummation, which is not performed here. However, Fig.~6 of Ref.~\cite{Liu:2018ktv} shows the ratio between the fixed-order NLO differential cross section and the one including only jet-radius resummation. A $\sim-10\%$ resummation effect is found at $\pt\sim100$ GeV, with smaller effects at higher transverse momenta. This is in good qualitative agreement with our findings for $R=0.4$, cf.~Fig.~\ref{fig:abs-pQCD-1}.

Interestingly, Ref.~\cite{Moch:2018hgy} found that NLO results, with or without resummation, are in good agreement with data, while the inclusion of NNLO corrections significantly worsens the agreement. This is at odds with our observation for the same jet radius $R=0.7$ that the NNLO corrections barely shift the central values and thus cannot worsen the agreement with the data. Indeed, we will not find such a worsening in the next section. The authors of Ref.~\cite{Moch:2018hgy} conclude that the large NNLO corrections they observe are a result of their scale choice. This is consistent with what was found in Ref.~\cite{Currie:2017ctp}. In that work, the inclusion of NNLO corrections likewise worsened the agreement with data if the central scales were chosen to be the leading jet $\pt$. When instead the central scale was set to the $\pt$ of the binned jet, small NNLO corrections which improve the description of the data were found. Combining the ratios between predictions at different orders and using different scales published in the works discussed here, we find that our NLO, NLO+NLL, and NNLO results are in good qualitative agreement with the literature.

\begin{figure}[t]
    \centering
    \includegraphics[width=0.5\linewidth]{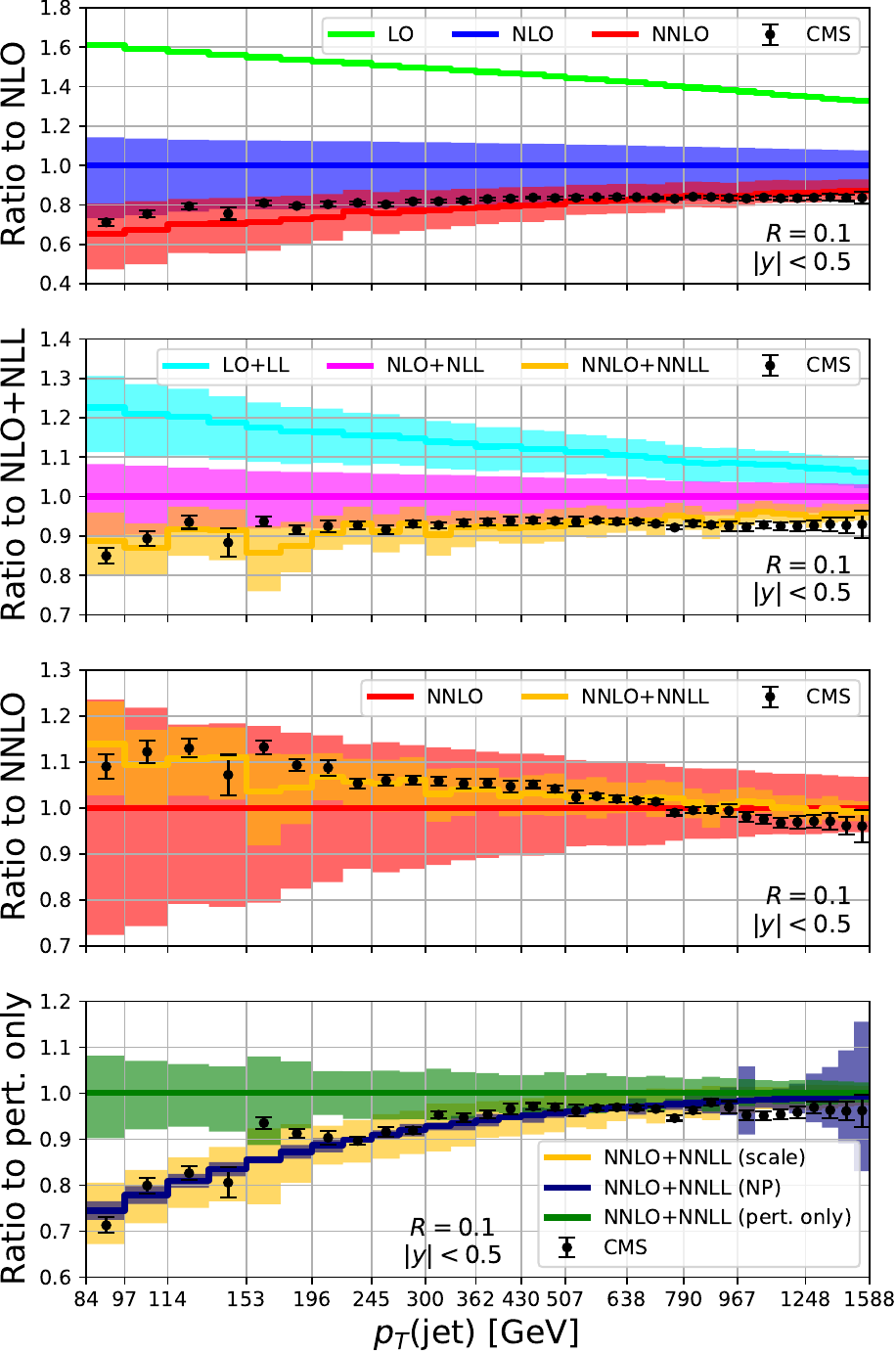}%
    \includegraphics[width=0.5\linewidth]{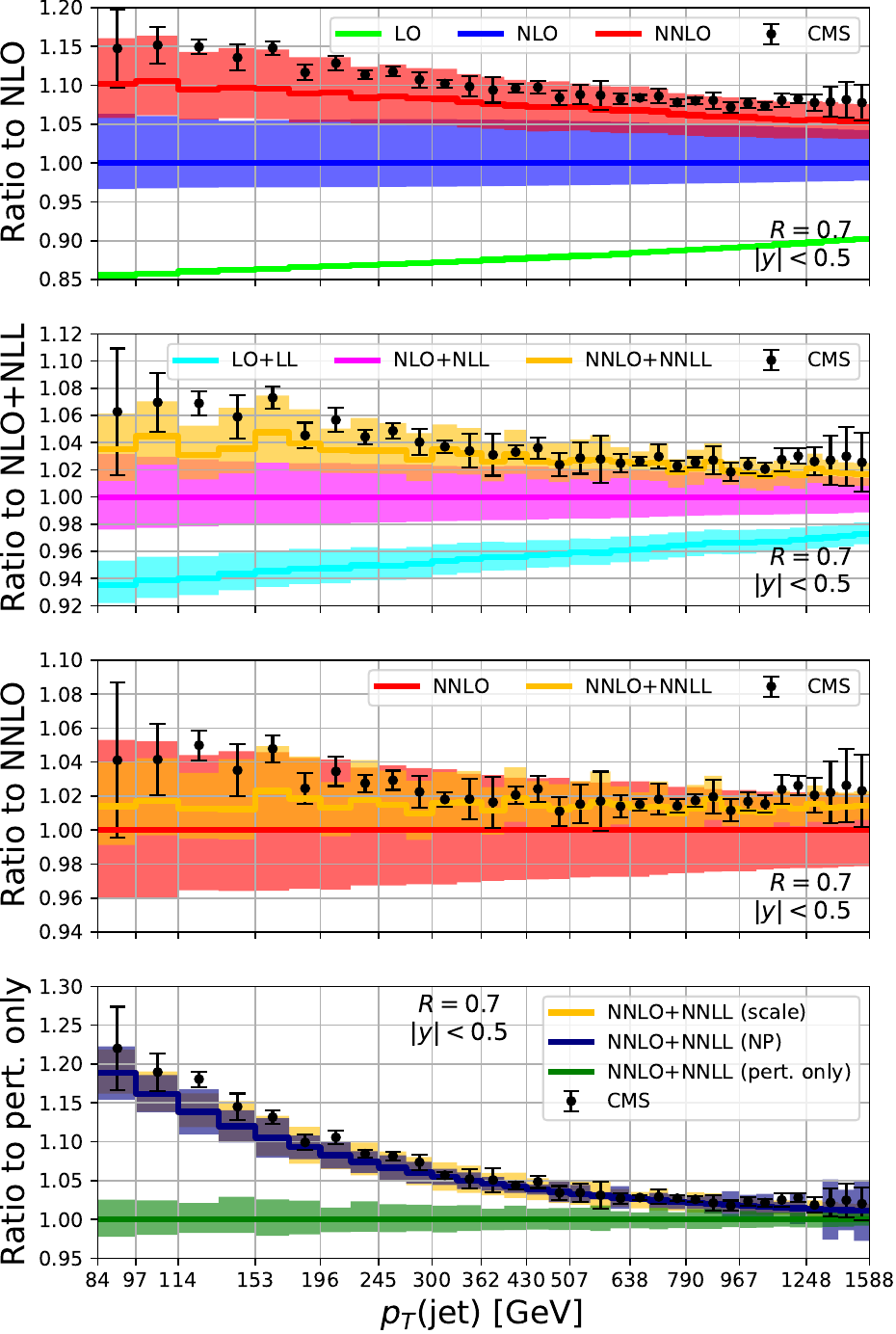}
    \caption{Comparison of the ratio $\mathrm{R}^{0.4}$ measured by CMS (black dots with errorbars) with perturbative and resummed predictions. The radius $R=0.1$ is on the left and $R=0.7$ on the right. The top panels show the fixed-order perturbative predictions relative to NLO, while the upper-middle panels show the resummed predictions relative to NLO+NLL. The lower-middle panels show the ratio with respect to NNLO. The colour coding for these is the same as in figure \ref{fig:abs-pQCD-1}. Finally, the bottom panels illustrate the non-perturbative corrections. In these panels, the strictly perturbative results are shown in green, while the corrected numbers are shown in yellow and blue, indicating the scale uncertainties and the uncertainties on the non-perturbative corrections, respectively.}
    \label{fig:ratio-cms-selection-1}
\end{figure}

\begin{figure}[t]
    \centering
    \includegraphics[width=0.5\linewidth]{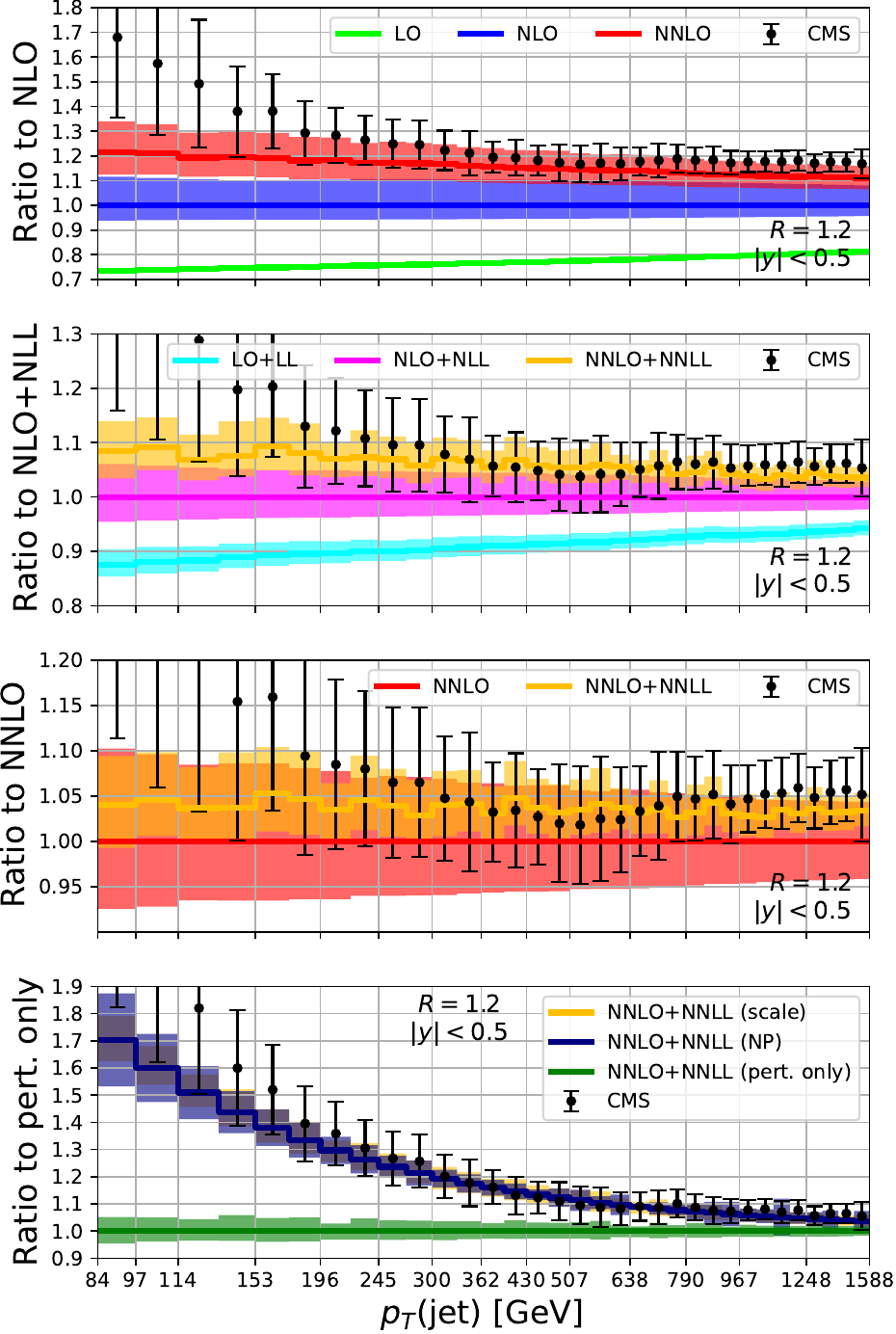}
    \caption{Same as \ref{fig:ratio-cms-selection-1} but for $R=1.2$.}
    \label{fig:ratio-cms-selection-2}
\end{figure}

\subsection{Comparison to CMS data}\label{subsec:cms}

We now turn to the comparison of the predictions with LHC data. The measurements by CMS in \cite{CMS:2020caw} are performed as ratios to the $R=0.4$ cross sections:
\begin{align}
    \mathrm{R}^{0.4}(R,\pt,\y) \equiv \frac{\dd \sigma(R) / \dd \pt/ \dd \y}{\dd \sigma(R=0.4) / \dd \pt / \dd \y}\;.
\end{align}
This in particular implies that $\mathrm{R}^{0.4}(0.4,\pt,\y) = 1$. The ratio is subject to non-perturbative corrections from hadronization and the underlying event. All fixed-order perturbative and resummed predictions we show in the following are corrected by the multiplicative factors we extract from the data provided by CMS in Ref.~\cite{CMS:2020caw}. A detailed discussion of the treatment of the non-perturbative corrections can be found in the Appendix~\ref{sec:non-perturbative}. The scale dependence used to estimate the MHOU for the ratio is computed fully correlated, i.e. the ratio is evaluated with the same scale choice in the numerator and the denominator. The same prescription is used for the different perturbative orders, i.e. the ratio is not expanded in $\alpha_s$, but the numerator and denominator are evaluated in the same perturbative order in $\alpha_s$.

\begin{figure}[t]
    \centering
    \includegraphics[width=0.5\linewidth]{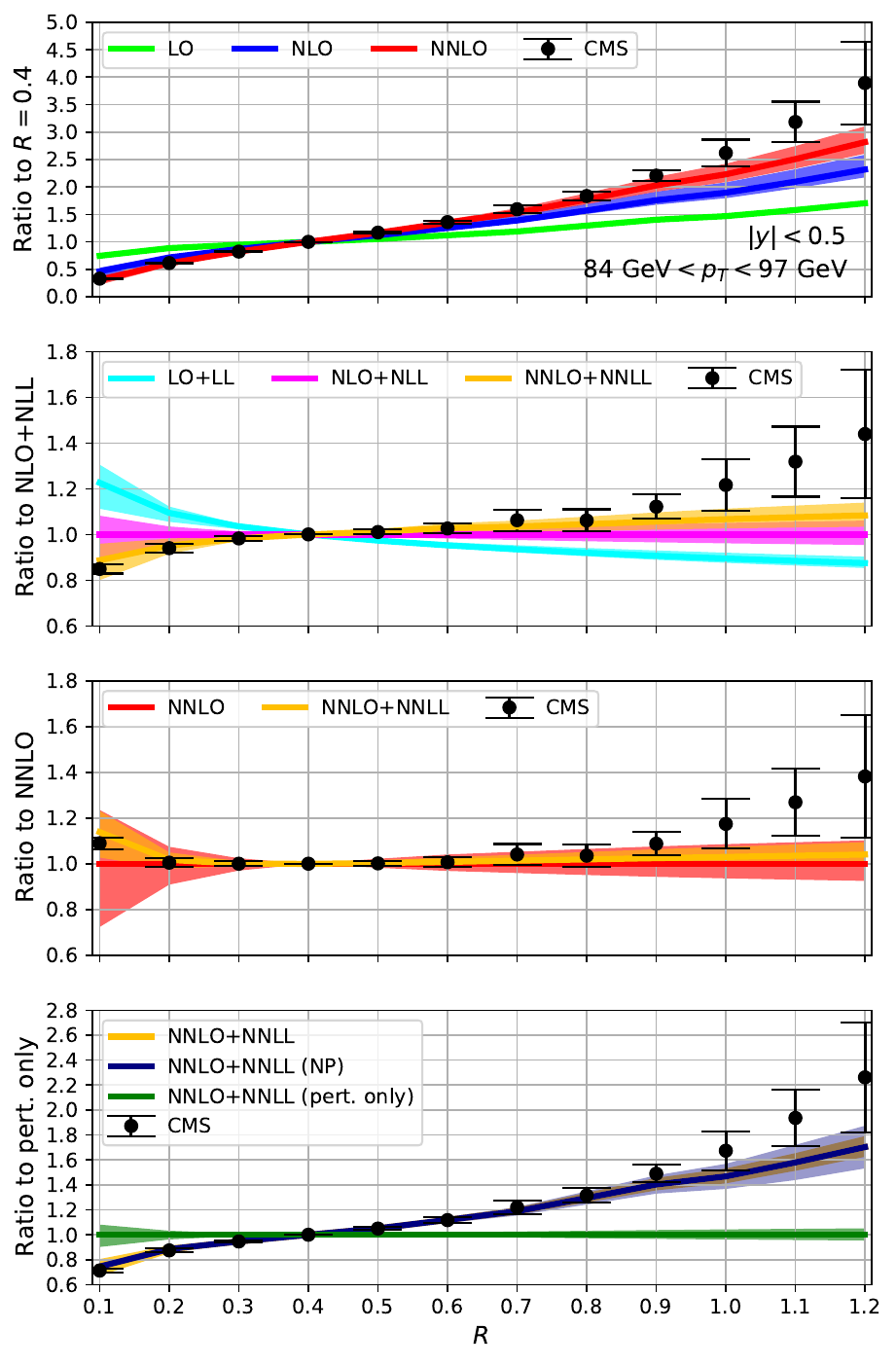}%
    \includegraphics[width=0.5\linewidth]{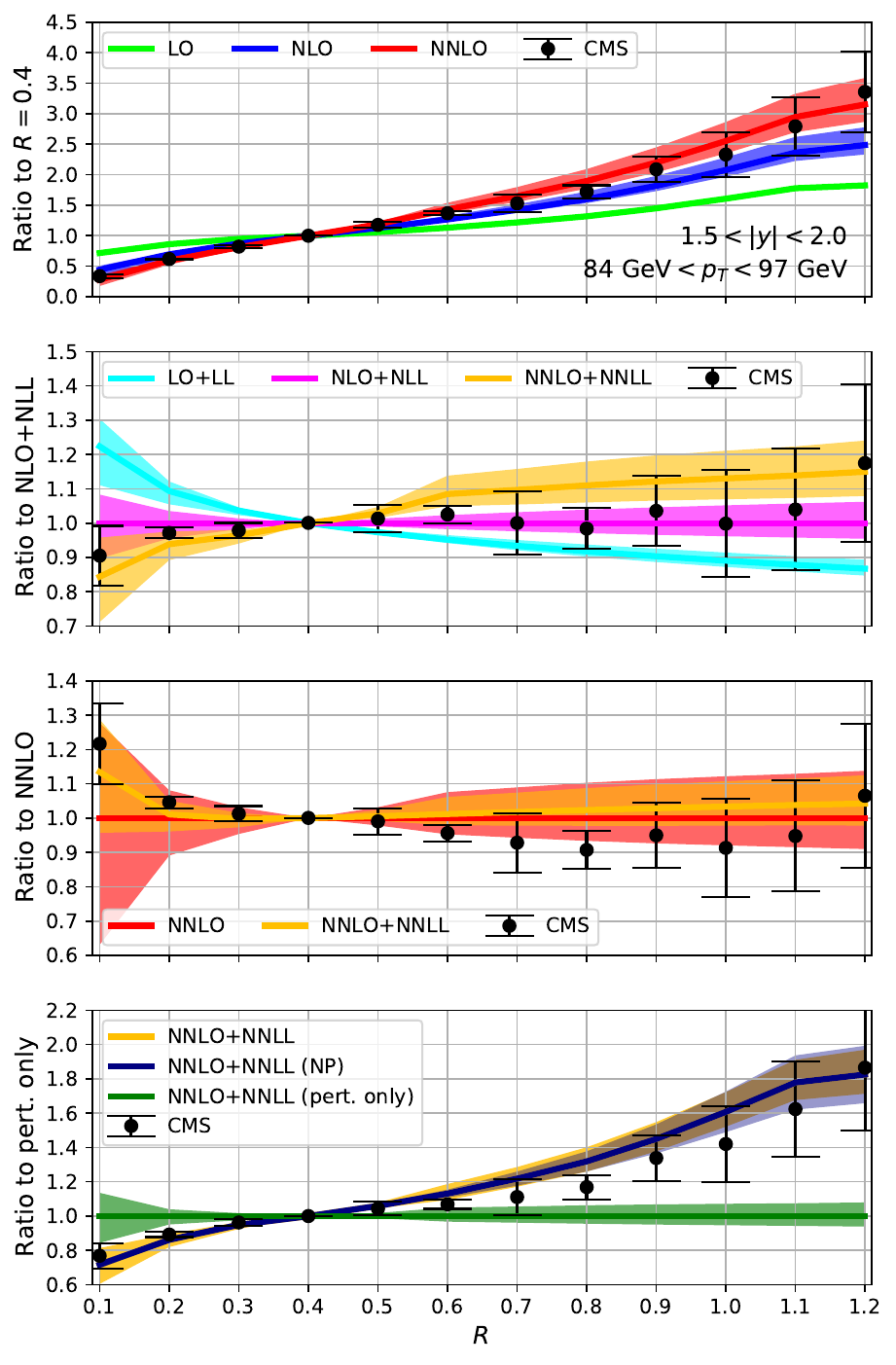}
    \caption{The $R$ dependence of $\mathrm{R}^{0.4}$ for $|\y| < 0.5$ (left) and $1.5 \leq |\y| < 2$ (right) for the $\pt\in[84,97)$ bin. The top panels show $\mathrm{R}^{0.4}(R)$ compared to fixed-order perturbative predictions. The second panels show the same but as ratio with respect to NLO+NLL compared to all resummed predictions. The third panels show a direct comparison between NNLO and NNLO+NNLL. The bottom panels show the size and uncertainty of the non-perturbative correction factors.}
    \label{fig:cms-r-dependence-1}
\end{figure}

We show the measured and predicted ratio $\mathrm{R}^{0.4}(R,\pt,\y)$ for $R=0.1,0.7$ and $1.2$ in Figs.~\ref{fig:ratio-cms-selection-1} and \ref{fig:ratio-cms-selection-2}. The data is only poorly described at fixed-order NLO QCD and is often outside the estimated uncertainties. Including NLL resummation helps overcome these discrepancies. In particular, for small radii, we see an improvement in the description already discussed in the literature \cite{Dasgupta:2016bnd}. Including second-order perturbative corrections significantly improves the description of the data. In most regions of the phase space, the data can be described with NNLO QCD within the uncertainties. Including the NNLL resummation further improves the picture, and across the spectrum, the central predictions follow the data. The improvement in shape can be particularly well gauged in the second-lowest panel, where a direct comparison between the NNLO and NNLO+NNLL predictions is shown. In addition to a better description of the shape, the scale dependence is reduced for the ratio when resummation is included. This indicates that the sensitivity of the spectrum to the scale is less dependent on $R$, as expected from the resummation, and thus cancels in the ratio to $R=0.4$. Finally, we indicate the impact of the non-perturbative effects in the lowest panel. We show NNLO+NNLO with scale variation with and without non-perturbative corrections and their associated uncertainties. We can see that for small $\pt$ the non-perturbative corrections are sizeable (negative for small $R$ and positive for large $R$) and, as expected, vanish for large $\pt$. The uncertainties on the corrections are small compared to the scale dependence of the NNLO+NNLL result for small radii $R$ except for the large $\pt$ region\footnote{The increase of the uncertainty is likely to be caused by the method to extract these corrections from Monte Carlo applied in Ref.~\cite{CMS:2020caw}, which suffers from insufficient statistics in these phase space regions.}. For larger radii, the sensitivity to non-perturbative effects increases which leads to uncertainties that are comparable to the remaining perturbative uncertainties.

\begin{figure}
    \centering
    \includegraphics[width=0.5\linewidth]{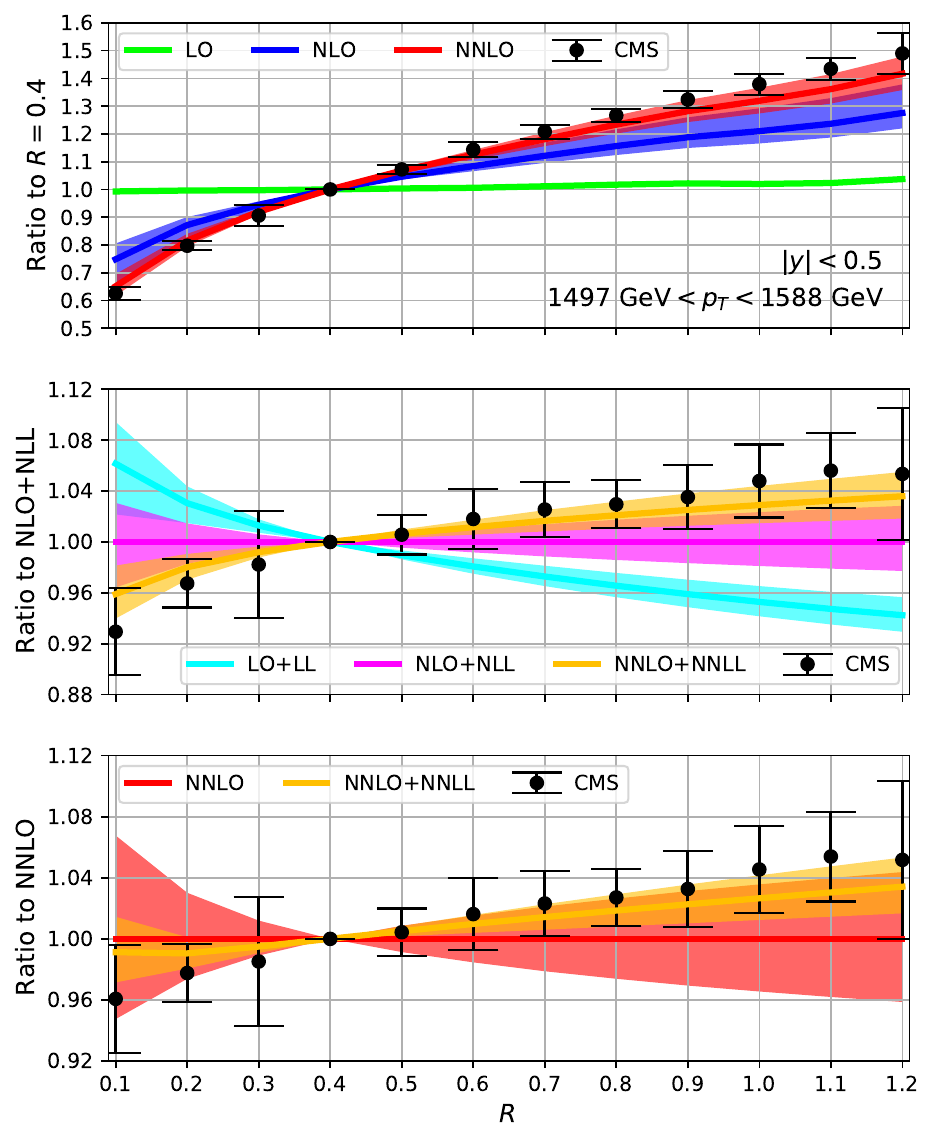}%
    \includegraphics[width=0.5\linewidth]{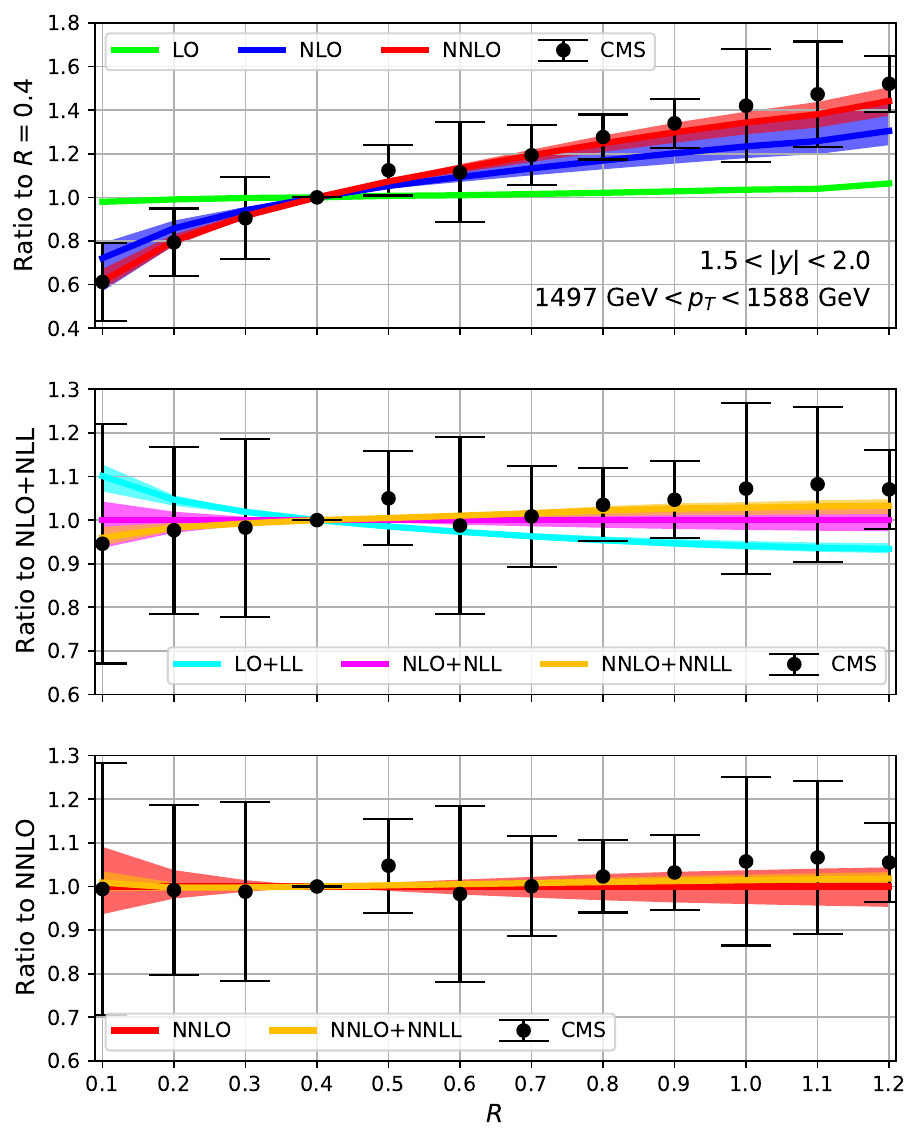}
    \caption{Same as \ref{fig:cms-r-dependence-1} but for  $\pt\in[1497,1588)$.}
    \label{fig:cms-r-dependence-2}
\end{figure}

In Fig.~\ref{fig:cms-r-dependence-1} and Fig.~\ref{fig:cms-r-dependence-2} we show the $R$ dependence of the ratio $\mathrm{R}^{0.4}$ for selected rapidity and transverse momentum bins, which represent the four corners of the 2D histogram. Again, we can see the improved description of the data when we include second-order corrections. The impact of the resummation can be seen clearly in the lower panels, where a direct comparison between NNLO and NNLO+NNLL is shown. The resummation generally reduces the scale dependence and moderately improves the shape. Overall, the data are very well described.

\section{Conclusion}\label{sec:conc}

In this paper, we have shown how to combine factorization theorems for small-$R$ jet production with fixed-order subtraction schemes for fragmentation, in particular the \textsc{Stripper} formalism. Combining recent developments in both areas, we have been able to achieve NNLO+NNLL predictions for small-$R$ inclusive jet production at hadron colliders. This is the first time NNLL accuracy has been achieved for small-$R$ jet resummation.

We studied a number of aspects of the factorization framework, in particular, power corrections in the jet radius. We found that these were small even for moderately sized jets, suggesting that this approach will enable precision predictions for a wide variety of jet-substructure observables.

Using our formalism, we presented phenomenological results for inclusive jet production at the LHC, and compared them with CMS data. We found that the inclusion of higher-order resummation greatly improved the agreement between theory and data at small-$R$, and stabilized the convergence and uncertainties from scale variations at larger $R$. It would be interesting to apply these predictions to precision extractions of $\alpha_s$ or PDFs, from inclusive jet data. To further improve the precision of our results, one direction in which our analysis could be extended is through the inclusion of threshold resummation. This has been studied in Refs.~\cite{Liu:2017pbb,Liu:2018ktv,Moch:2018hgy}, it would be interesting to incorporate it in our framework and study its phenomenological implications.

While the primary focus of this paper was on inclusive jet production, our method extends to any jet substructure observables that exhibit a factorization theorem onto an inclusive hard function. This includes, in particular, energy correlator observables \cite{Basham:1978zq,Basham:1979gh,Basham:1977iq,Basham:1978bw}. Following their introduction as a jet substructure observable \cite{Chen:2020vvp,Komiske:2022enw} they have recently been measured in a number of hadronic colliders, including proton-proton \cite{STAR:2025jut,Komiske:2022enw,CMS:2024mlf,ALICE:2024dfl}, proton-nucleus \cite{talk_Anjali}, and nucleus-nucleus \cite{CMS-PAS-HIN-23-004} collisions. They can be described by an inclusive factorization theorem \cite{Lee:2022ige,Lee:2024tzc,Lee:2024icn}, which builds on Ref.~\cite{Dixon:2019uzg} by incorporating the identification of the hadronic jet. Most excitingly, Ref.~\cite{CMS:2024mlf} was able to perform a precision extraction of the strong coupling constant from the energy correlators. Since the energy correlator jet functions are known to two-loops \cite{Dixon:2019uzg,Chen:2023zlx}, the primary theoretical uncertainty comes from the description of the inclusive jet production, which we have addressed in this paper. We look forward to extending the approach of this paper to the calculation of the energy correlators in future work, which we hope will enable a precision era of jet substructure at hadron colliders.

\section*{Acknowledgements}
T.G.~would like to thank Sven-Olaf Moch for interesting discussions on past calculations. K.L. was supported by the U.S. Department of Energy, Office of Science, Office of Nuclear Physics from DE-SC0011090. I.M. is supported by the DOE Early Career Award DE-SC0025581. XY.Z. was supported in part by the U.S. Department of Energy under contract DE-SC0013607. T.G.~has been supported by STFC consolidated HEP theory grants ST/T000694/1 and ST/X000664/1. This work was performed using the Cambridge Service for Data Driven Discovery (CSD3), part of which is operated by the University of Cambridge Research Computing on behalf of the STFC DiRAC HPC Facility (www.dirac.ac.uk). The DiRAC component of CSD3 was supported by STFC grants ST/P002307/1, ST/R002452/1 and ST/R00689X/1.

\clearpage 

\appendix

\section{Perturbative power corrections}
\label{sec:power-corrections}

\begin{figure}[h]
    \centering
    \includegraphics[width=0.5\linewidth]{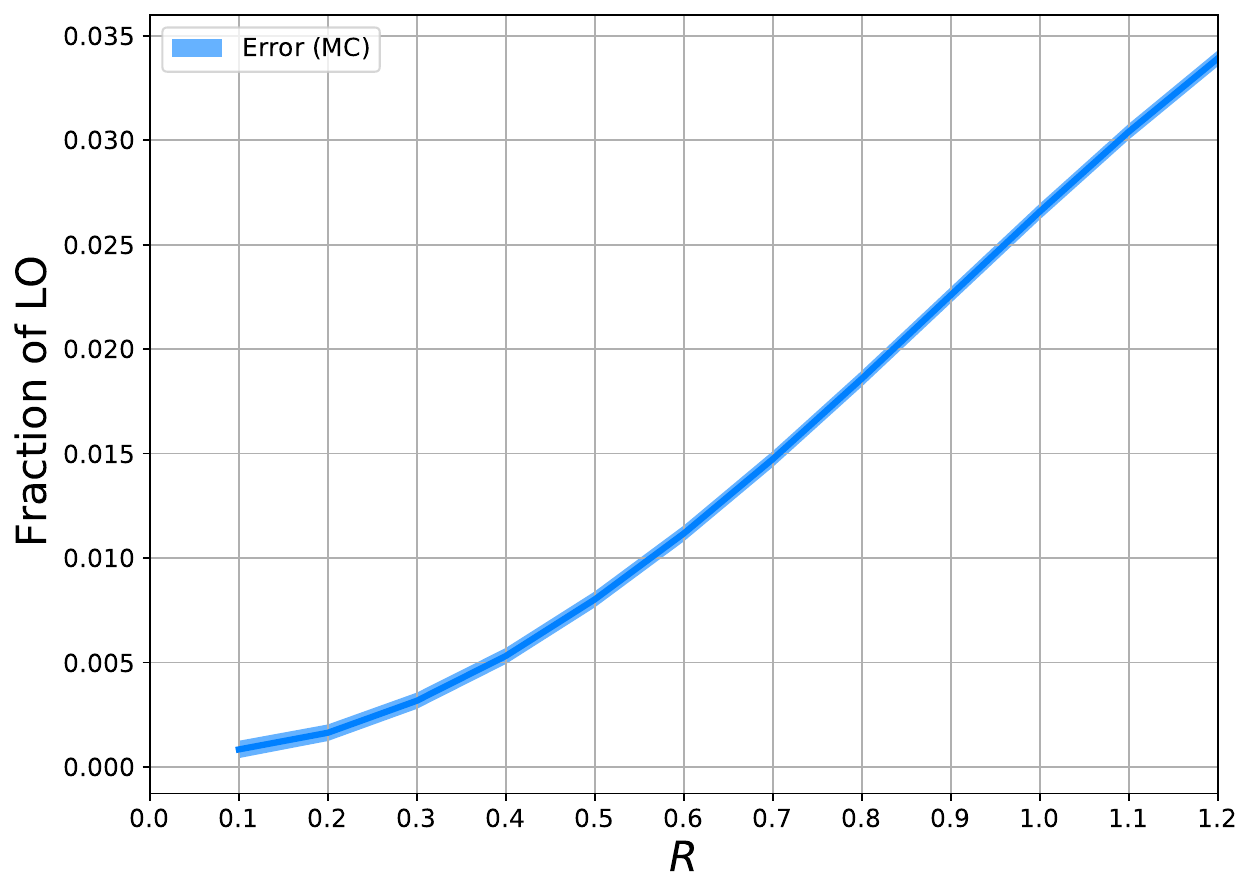}%
    \includegraphics[width=0.5\linewidth]{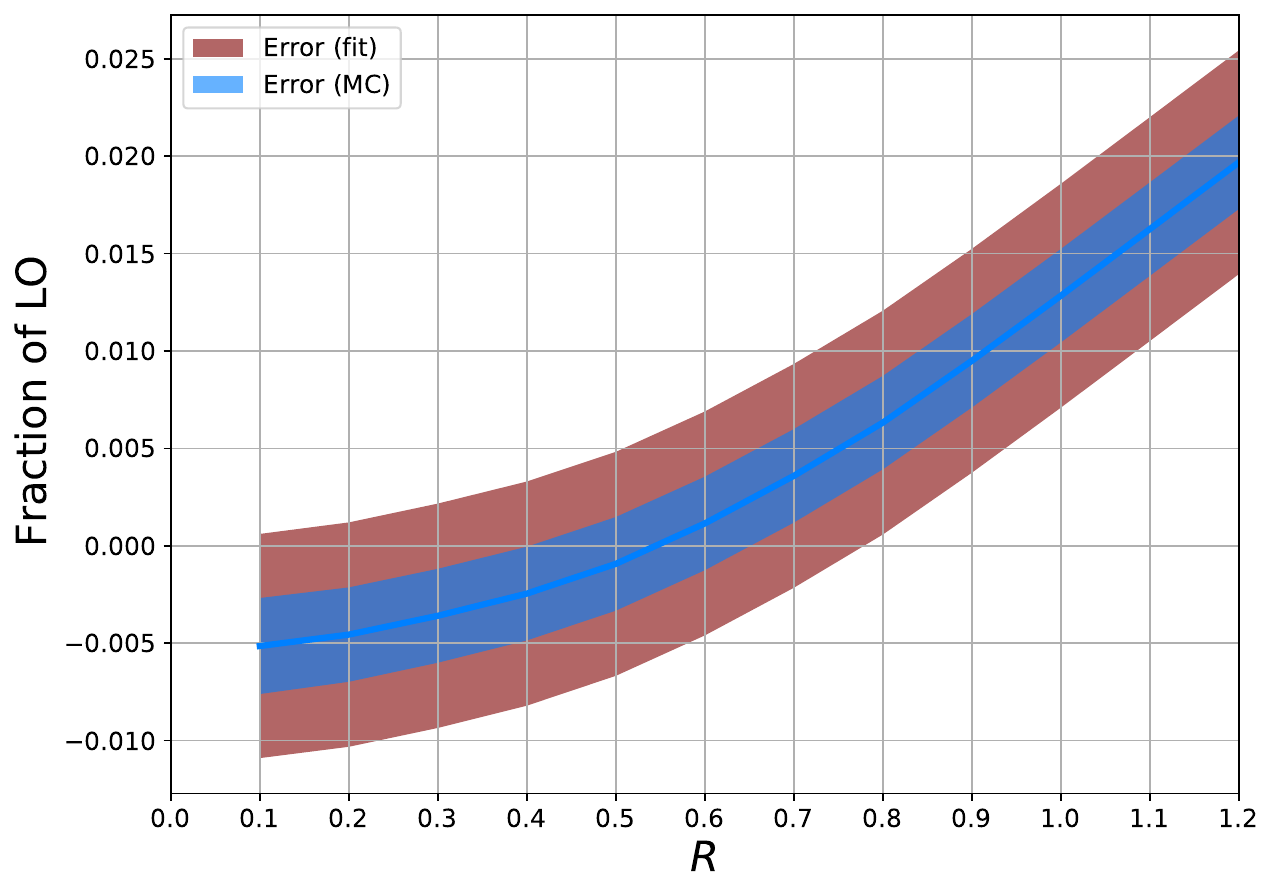}
    \caption{The size of the power corrections to the NLO (left) and NNLO (right) corrections, as a fraction of the LO cross section. The blue error band shows the MC error, while the dark red band shows the propagated uncertainties on the NNLO jet function constants.}
    \label{fig:PowerCorr}
\end{figure}

The factorization theorem in Eq.~\eqref{eq:factorization} is subject to quadratic perturbative power corrections in the jet-radius parameter $R$. We can test these by comparing fixed order computations, which contain all power corrections at a given order in $\alpha_s$, with the expanded factorization formula at the same order. This comparison is shown in Fig.~\ref{fig:PowerCorr} for NLO and NNLO QCD, where we show the corrections as a fraction of the LO cross section. More specifically, these plots show the weighted averages over all $\pt$ bins for $|\y| < 0.5$. This is done to significantly reduce the Monte Carlo error. It was checked that the corresponding plot for each of the 33 bins is consistent with this average. The expected quadratic behavior in $R$ is observed and the power corrections at small $R$, i.e. $R=0.1$, are of per mille level. The small size of the corrections justifies the extraction of the NNLO jet constant as performed in Sec.~\ref{sec:framework} at small $R$. It also motivates us to think that such a factorization formula can be used in similar scenarios such as small-angle energy correlators and other jet-substructure observables.

\section{Treatment of non-perturbative corrections}
\label{sec:non-perturbative}

\begin{figure}
    \centering
    \includegraphics[width=0.5\linewidth]{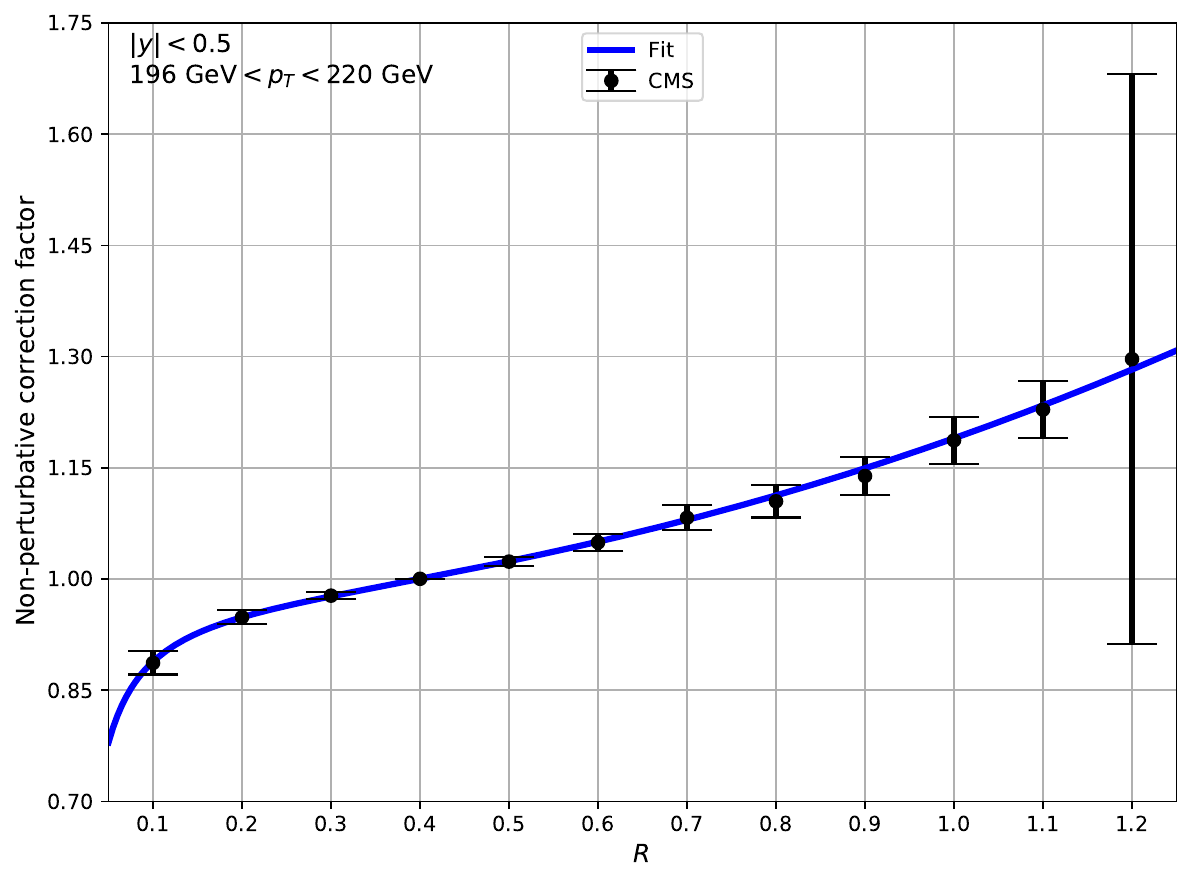}%
    \includegraphics[width=0.5\linewidth]{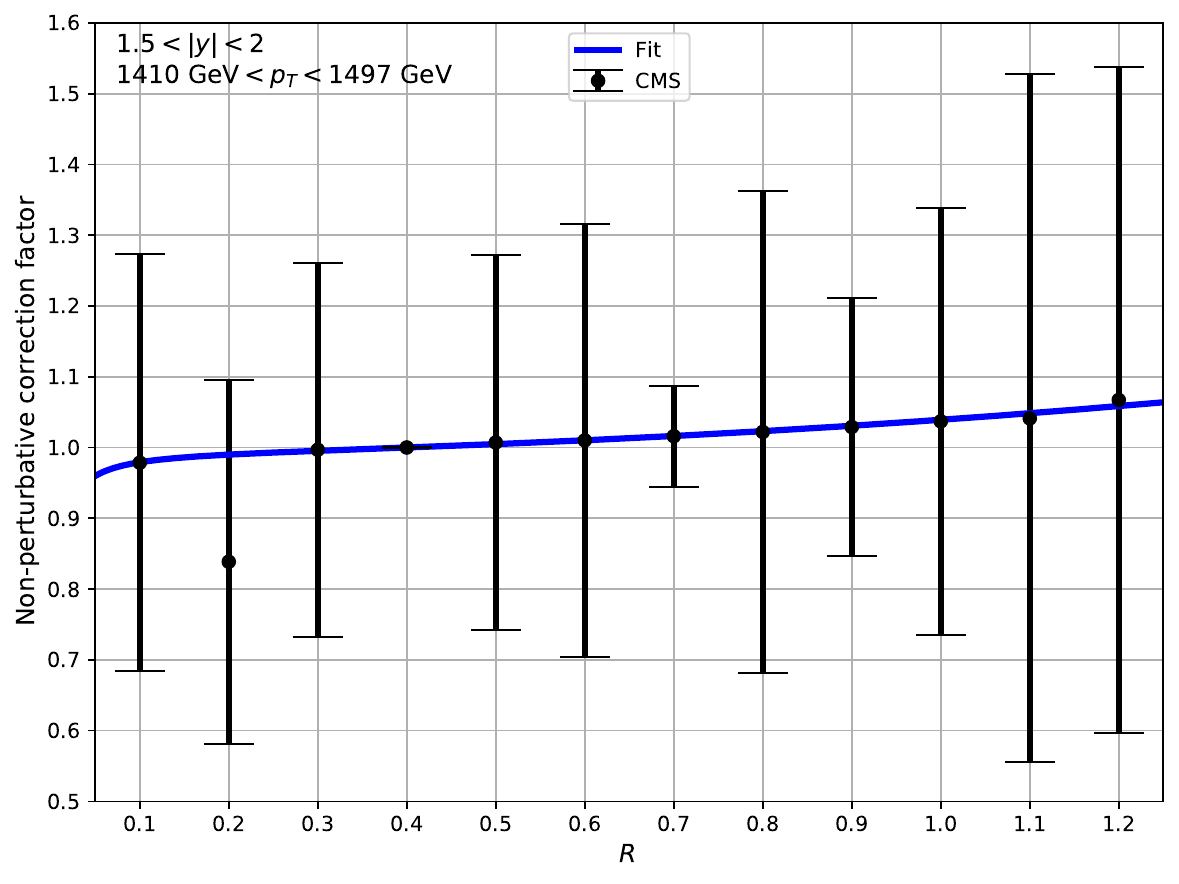}
    \caption{Non-perturbative corrections for two different $\pt$ and $\y$ bins provided by CMS including fits to analytic forms as discussed in the text.}
    \label{fig:app:np_fits}
\end{figure}

The perturbative predictions require corrections to compensate for non-perturbative effects from hadronization and underlying event before they can be compared to data. In Sec.~\ref{sec:phenomenology} we used the correction factors provided by CMS~\cite{CMS:2020caw}, which have been computed using a Monte Carlo simulation with \textsc{Powheg} + \textsc{Pythia} and \textsc{Powheg} + \textsc{Herwig++}. The provided corrections contain a few artifacts which are (likely) of a numerical or statistical nature, and we corrected them as follows:
\begin{enumerate}
    \item  The provided errors for $R=1.2$ appear to be wrong (far too large), see LHS of Fig.~\ref{fig:app:np_fits}. We estimate them by taking the errors for $R=1.1$ and rescaling by $(k_{\text{non-pert.}}(R=1.2) - 1) /(k_{\text{non-pert.}}(R=1.1) - 1)$. This is based on the idea that the uncertainty should stem from the uncertainty on $\Lambda_\text{UE}$ and that $( k_{\text{non.pert}}(R \gg 1) - 1)$ should be proportional to $\Lambda_\text{UE}$.
    \item The central values for $R=0.2$ for the rapidity slice $1.5<|\y|<2.0$ also appear to be incorrect and have been replaced with values obtained from a fit to all other values of $R$ of the form $(1+A/R+B\:R^2)/(1+A/0.4+B\:0.4^2)$, which is repeated independently for each $\pt$ bin, see RHS of Fig.~\ref{fig:app:np_fits}.
\end{enumerate}

These corrections are motivated by the known scaling properties of these corrections, both in terms of $\pt$ and in terms of $R$. Ref.~\cite{Dasgupta:2007wa} considered linear power corrections (i.e.~$\mathcal{O}(\Lambda_\text{QCD}/\pt)$) and found that hadronization corrections scale like $1/R$, while underlying event contributions scale like $R^2$. As a cross-check of the non-perturbative corrections provided in the CMS paper, we can verify that these scalings are present. Fig.~\ref{fig:app:np_fits} shows that a model of the type
\begin{align}
    \frac{\dd \sigma(R)}{\dd \sigma^{\text{pert.}}(R)} = 1+A/R+B\:R^2
    \label{eq:model-np-1}
\end{align}
can describe the power corrections very well. Although this is shown here for only two $\pt$ bins, this has been verified for all $\pt$ bins. Therefore, the dependence $R$ is consistent with the theoretical expectation.

One can also perform a 2D fit to the full $R$ and $\pt$ dependence, using Eq.~(5.9) of Ref.~\cite{Dasgupta:2007wa}:
\begin{align}
    \frac{\dd \sigma}{\dd \pt} (\pt) =
    \frac{\dd \sigma^{\text{pert.}}_q}{\dd \pt} \left(1+n_q \frac{\langle \delta \pt^q \rangle_{\text{NP}}}{\pt}\right) + \frac{\dd \sigma^{\text{pert.}}_g}{\dd \pt} \left(1+n_g \frac{\langle \delta \pt^g \rangle_{\text{NP}}}{\pt}\right)\,.
    \label{eq:model-np-2}
\end{align}
The parameters $n_i$ correspond to the power of the transverse momentum spectrum for each parton species (here quarks $q$ and gluons $g$), i.e. $\pt^{-n_i}$, and the (species-dependent) non-perturbative average transverse momentum shifts are:
\begin{align}
    \langle \delta \pt^i \rangle_{\text{NP}} = - 2 \frac{C_i}{R}\mathcal{A}(\mu_I) + R J_1(R) \Lambda_{\text{UE}}\;,
\end{align}
with the first moment of the non-perturbative $\alpha_s$ denoted by $\mathcal{A}(\mu_I)$. Our 2D fit yields values of 0.23 GeV and 14 GeV for $2C_F\mathcal{A}$ and $\Lambda_\text{UE}$, respectively, qualitatively in agreement with the values of $\sim$0.5 GeV and $\sim$10 GeV given in Ref.~\cite{Dasgupta:2007wa}. A better fit is obtained by absorbing a factor $f_qn_q+f_gn_g$ into the definitions of the non-perturbative scales. Either way, we can verify that the non-perturbative corrections provided by CMS qualitatively behave as expected from theory.


\bibliographystyle{JHEPmod}
\bibliography{lit}

\end{document}